\documentclass[a4paper,11pt]{article}
\pdfoutput=1 

\usepackage{jcappub}
\usepackage{dsfont}
\usepackage{dcolumn}

\newcommand{\tr}{\mathop{\mathrm{tr}}}

\title{Dynamical system approach to running $\Lambda$ cosmological models}

\author[a]{Aleksander Stachowski}
\author[a,b]{Marek Szyd{\l}owski}

\affiliation[a]{Astronomical Observatory, Jagiellonian University, Orla 171, 30-244 Krakow, Poland}
\affiliation[b]{Mark Kac Complex Systems Research Centre, Jagiellonian University, {\L}ojasiewicza 11, 30-348 Krak{\'o}w, Poland}

\emailAdd{aleksander.stachowski@uj.edu.pl}
\emailAdd{marek.szydlowski@uj.edu.pl}

\abstract{
We discussed the dynamics of cosmological models in which the cosmological constant term is a time dependent function through the scale factor $a(t)$, Hubble function $H(t)$, Ricci scalar $R(t)$ and scalar field $\phi(t)$. We considered five classes of models; two non-covariant parametrization of $\Lambda$:
		1) $\Lambda(H)$CDM cosmologies where $H(t)$ is the Hubble parameter,
		2) $\Lambda(a)$CDM cosmologies where $a(t)$ is the scale factor, and three covariant parametrization of $\Lambda$:
		3) $\Lambda(R)$CDM cosmologies, where $R(t)$ is the Ricci scalar,
		4) $\Lambda(\phi)$-cosmologies with diffusion,
		5) $\Lambda(X)$-cosmologies, where $X=\frac{1}{2}g^{\alpha\beta}\nabla_{\alpha}\nabla_{\beta}\phi$ is a kinetic part of density of the scalar field.
We also considered the case of an emergent $\Lambda(a)$ relation obtained from the behavior of trajectories in a neighborhood of an invariant submanifold. In study of dynamics we use dynamical system methods for investigating how a evolutional scenario can depend on the choice of special initial conditions. We showed that methods of dynamical systems offer the possibility of investigation all admissible solutions of a running $\Lambda$ cosmology for all initial conditions, their stability, asymptotic states as well as a nature of the evolution in the early universe (singularity or bounce) and a long term behavior at the large times. We also formulated an idea of the emergent cosmological term derived directly from an approximation of exact dynamics. We show that some non-covariant parametrizations of Lambda term like $\Lambda(a)$, $\Lambda(H)$ give rise to pathological and nonphysical behaviour of trajectories in the phase space. This behaviour disappears if the term $\Lambda(a)$ is emergent from the covariant parametrization.
}

\begin{document}
\maketitle
\flushbottom

\section{Introduction}
Our understanding of properties of the current Universe based on the assumption that gravitational interactions, which are extrapolating at the cosmological scales, are described successfully by the Einstein general relativity theory with the cosmological term $\Lambda$. If we assume that the geometry of the Universe is described by the Robertson-Walker metric, i.e., the universe is spatially homogeneous and isotropic then we obtain the model of the current Universe in the form of standard cosmological model (the $\Lambda$CDM model). From the methodological point of view this model plays the role of a kind of the effective theory which describes well the current Universe in the present accelerating epoch.

If we compare the $\Lambda$CDM model with the observational data then we obtain that more than $70\%$ of matter content is in the form of dark energy and well modeled in terms of an effective parameter of the cosmological constant term.

In order if we assume that the SCM (standard cosmological model) is an EFT (effective field theory) which is valid up to a certain cutoff of mass $M$, then Weinberg's argument based on naturalness is that $M_{\text{pl}}^2 \Lambda \propto M^4$ \cite{Weinberg:1988cp}. This means that if we extrapolate of the SCM up to the Planck scale then we should have $\Lambda\propto M_{\text{pl}}^2$. On the other hand from the observation we obtain that both density parameters $\Omega_{\Lambda,0}=\frac{\Lambda}{3H_0^2}$ and $\Omega_{\text{m},0}=\frac{\rho_{\text{m},0}}{3H_0^2}$ are order one, which implies $\Lambda\propto H_0^2 \propto  M_{pl}^2 \times 10^{-120}$.

In consequence we obtain the huge discrepancy between the expected and observed values of the $\Lambda$. It is just what is called the cosmological constant problem requiring the explanation why the cosmological constant assumes such a small value today.

In this context an idea of a running cosmological constant term appears. Shapiro and Sola \cite{Shapiro:2009dh} showed neither there is the rigorous proof indicating that the cosmological constant is running, nor there are strong arguments for a non-running one. Therefore one can study different theoretical possibilities of the running $\Lambda$ term given in a phenomenological form and investigate cosmological implications of such an assumption.

The corresponding form of the $\Lambda(t)$ dependence can be motivated by quantum field theory~\cite{Shapiro:2009dh,Bonanno:2011yx,Urbanowski:2014gza} or by some theoretical motivations~\cite{Lima:2014hia,Alcaniz:2005dg}. Padmanabhan~\cite{Padmanabhan:2001pp,Vishwakarma:2002eka} suggested also that $\Lambda \propto H^2$ from the dimensional considerations.

The main aim of this paper is to study dynamics of the cosmological models of the $\Lambda(t)$CDM cosmologies in which matter is given in the form of dust. Such models are a simple generalization of the standard cosmological model in which the $\Lambda$ is constant. The relation $\Lambda(t)$ is not given directly but through the function which describes the evolution of the Universe.

It is considered two classes of the models with the non-covariant parametrizations of the $\Lambda$ term:
\begin{itemize}
	\item the cosmological models in which dependence on time is hidden and $\Lambda(t)=\Lambda(H(t))$ or $\Lambda(t)=\Lambda(a(t))$ depends on the time through the Hubble parameter $H(t)$ or scale factor $a(t)$,
\end{itemize}
and three classes of the models with covariant parametrizations of the $\Lambda$ term:
\begin{itemize}
	\item the Ricci scalar of the dark energy model, i.e., $\Lambda=\Lambda(R)$,
	\item the parameterization of the $\Lambda$ term through the scalar field $\phi(t)$ with a self-interacting potential $V(\phi)$,
	\item  as the special case of the previous one, the $\Lambda$ term can be parametrized by a kinetic part of the energy density of the scalar field $X=\frac{1}{2}g^{\alpha\beta}\nabla_\alpha \nabla_\beta\phi$.
\end{itemize}

Note that some parametrizations of the $\Lambda$ term can also arise from another theory beyond general relativity. For example Shapiro and Sola \cite{Shapiro:2009dh} suggested that a solution derived the form of $\rho_{\Lambda}(H)=\rho_{\Lambda}^0+\alpha(H^2-H_0^2)+\mathcal{O}(H^4)$ as a solution of fundamental general relativity equations.

Dynamics of both subclasses of $\Lambda(t)$CDM cosmologies is investigated by dynamical system methods. The main advantage of these methods is a possibility of study all solutions with admissible initial conditions. In this approach we are looking for attractor solutions in the phase space representing generic solutions for the problem which gives such a parameterization of $\Lambda(t)$ which explain how the value of cosmological term achieves a small value for the current Universe. Bonanno and Carloni have recently used the dynamical sustems approach to study the qualitative behaviour of FRW cosmologies with time-dependent vacuum energy on cosmological scales \cite{Bonanno:2011yx}.

We also develop an idea of emergent relation $\Lambda(a)$ obtained from behaviour of trajectories of the dynamical system near the invariant submanifold $\frac{\dot{H}}{H^2}=0$. By emerging running parametrization $\Lambda(a)$ we understand its derivation directly from the true dynamics. Therefore, corresponding parametrization is obtained is obtained from the entry of trajectories in a de Sitter state.

\section{$\Lambda(H)$CDM cosmologies as a dynamical systems}
From the theoretical point of view if we do not know an exact form of the $\Lambda(t)$ relation we study dynamical properties of cosmological models in which dependence of the $\Lambda$ on $t$ is through the Hubble parameter or scale factor, i.e. $\Lambda(t)=\Lambda(H(t))$ or $\Lambda(t)=\Lambda(a(t))$. It will be demonstrated that the connection of such models with the mentioned in the previous section in which choice of $\Lambda(t)$ form was motivated by physics. Cosmological models with quadratic dependence of Lambda on cosmological time reveals as a special solutions in the phase space.

In investigation of dynamics of $\Lambda(H)$ cosmologies it would be useful to apply dynamical system methods~\cite{Perko:2001de}. The main advantage of these methods in the context of cosmology is that they offer the possibility of investigations of all solutions, which are admissible for all physically admitted initial conditions. The global characteristics of the dynamics is given in the form of phase portraits, which reflects the phase space structure of all solutions of the problem. Then one can distinguish some generic (typical) cases as well non-generic (fine tunings) which physical realizations require a tuning of initial conditions. The methods of dynamical systems gives us also possibility of the study stability of the solutions in a simply way by investigation of linearization of the system around the non-degenerated critical points of the system.

If the dynamical system is in the form $\dot{\mathbf{x}}\equiv\frac{d\mathbf{x}}{dt}=f(\mathbf{x})$, where $\mathbf{x}\in\mathbb{R}^n$ and $f$ is $C^{\infty}$ class, then solutions of this system is a vector field $\mathbf{x}(t;\mathbf{x}_0)$ where $\mathbf{x}(t_0)$ is the vector of initial conditions. Beyond this regular solution there are singular solutions. They are special solutions obtained from a condition of vanishing its right-hand sides.

The $\Lambda(H)$CDM cosmological models have been recently investigated intensively in the contemporary cosmology~\cite{Bessada:2013maa,Graef:2013iia,Lima:2014hia,Perico:2013mna}. Among these class of models cosmology with particular form of $\Lambda(t)=\Lambda+\alpha H^2$ is studied in details~\cite{Bessada:2013maa} as well as its generalization to the relation of $\Lambda(H)$ given in the form of a Taylor series of the Hubble parameter~\cite{Gomez-Valent:2014rxa}.

It is also interesting that motivations for study such a class of models can be taken from Urbanowski's expansion formula for decaying false vacuum energy which can be identified with the cosmological constant term~\cite{Urbanowski:2014gza}. It is sufficient to interpret time $t$ in terms of the Hubble time scale $t=t_H \equiv \frac{1}{H}$. Therefore $\Lambda(H)$CDM cosmologies can be understand as some kind of an effective theories of describing influence of vacuum decay in the universe~\cite{Wang:2004cp}. This approach is especially interesting in the context of both dark energy and dark matter problem because the problem of cosmological constant cannot be investigated in isolation to the problem of dark matter.

In $\Lambda(H)$ cosmologies, in general, a scaling relation on matter is modified and differs from the canonical relation $\rho_{\text{m=}}\rho_{\text{m},0}a^{-3}$ in the $\Lambda$CDM model. The deviation from the canonical relation here is characterized by a positive constant $\epsilon$ such that $\rho_{\text{m}}=\rho_{\text{m},0} a^{-3+\epsilon}$~\cite{Costa:2007sq}.

FRW cosmologies with a running cosmological term $\Lambda(t)$ such that $\rho_{\text{vac}}=\Lambda(t)$ and $p_{\text{vac}}=-\Lambda(t)$ can be formulated in the form of a system of non-autonomous dynamical system
\begin{align}
\frac{dH}{dt} &\equiv\dot H = -H^2-\frac{1}{6}(\rho_m+3p_m)+\frac{1}{3}\Lambda(t)\label{sys1} \\
\frac{d\rho_{\text{m}}}{dt} &\equiv\dot\rho_m = -3H(\rho_m+p_m)-\dot\Lambda,\label{sys2}
\end{align}
where $\rho_m$ and $p_m$ are energy density of matter and pressure, respectively, and a dot denotes differentiation with respect to the cosmological time. In this cosmology an energy-momentum tensor is not conserved because the presence of an interaction in both matter and dark energy sector. System (\ref{sys1})-(\ref{sys2}) has a first integral called the conservation condition in the form
\begin{equation}
\rho_{\text{m}}-3H^2=\Lambda(t).\label{integral}
\end{equation}
Note that a solution $\rho_{\text{m}}=0$ is a solution of (\ref{sys2}) only if $\Lambda=\text{const}$.
Of course (\ref{sys1})-(\ref{sys2}) does not forms the closed dynamical system while the concrete form of $\Lambda(t)$ relation is not postulated. Therefore this cosmology belongs to a more general class of models in which energy momentum tensor of matter is not conserved.

Let us consider that both visible matter and dark matter are given in the form of dust, i.e. $p_{\text{m}}=0$ and
\begin{equation}
\Lambda(t)=\Lambda(H(t)).\label{assumption}
\end{equation}
Due to the above simplifying assumption (\ref{assumption}) the system (\ref{sys1})-(\ref{sys2}) with the first integral in the form (\ref{integral}) assumes the form of a two-dimensional closed dynamical system
\begin{align}
\dot H &=-H^2-\frac{1}{6}\rho_{\text{m}}+\frac{1}{3}\Lambda(t),\label{sys3} \\
\dot\rho_{\text{m}} &=-3H\rho_{\text{m}}-\Lambda'(H)\left(-H^2-\frac{1}{6}\rho_{\text{m}}+\frac{\Lambda(H)}{3}\right),\label{sys4}
\end{align}
where $\Lambda'(H)=\frac{d\Lambda}{dH}$ and $\rho_m-3H^2=-\Lambda(H)$ is the first interval of the system (\ref{sys3})-(\ref{sys4}).

Let us consider $\Lambda(H)$ given in the form of a Taylor series with respect to the Hubble parameter $H$, i.e.
\begin{equation}
\Lambda(H)=\sum_{n=1}^\infty \frac{1}{n!}\frac{d^n}{dH^n} \Lambda(H)|_0 H^n.\label{series}
\end{equation}
We assume additionally that the model dynamics has a reflection symmetry $H\rightarrow -H$, i.e., $a(t)$ is a solution of the system and $a(-t)$ is also its solution. Therefore only even terms of type $H^{2n}$ are present in the expansion series (\ref{series}). Finally we assume the following form of energy density parametrization through the Hubble parameter $H$
\begin{equation}
\rho_\Lambda (H)=\Lambda_{\text{bare}}+\alpha_2 H^2+\alpha_4 H^4+\cdots .
\end{equation}
There are also some physical motivations for such a choice of $\Lambda(H)$ parametrization (see \cite{Perico:2013mna}).

It would be useful for a further dynamical analysis of the system under consideration to reparametrize the time variable
\begin{equation}
\tau\longmapsto\tau=\ln a
\end{equation}
and rewrite dynamical system (\ref{sys3})-(\ref{sys4}) in new variables:
\begin{equation}
x=H^2,\text{ }y=\rho_m.
\end{equation}
Then we obtain the following dynamical system
\begin{align}
x' &\equiv\frac{dx}{d\ln a}=2\left[-x-\frac{1}{6}y+\frac{1}{3}(\Lambda+\alpha_2 x+\alpha_4 x^2 + \cdots)\right],\label{dyn4p}\\
y' &\equiv\frac{dy}{d\ln a}=-3y-\frac{1}{3}(\alpha_2+2\alpha_4 x+\cdots)\left[-x-\frac{1}{6}y+\frac{1}{3}(\Lambda+\alpha_2 x+\alpha_4+\cdots)\right]\label{dyn5p}
\end{align}
and
\begin{equation}
y-3x=-(\Lambda+\alpha_2x+\alpha_4x^2+\cdots)\label{integral2}
\end{equation}
where instead of $\Lambda_{\text{bare}}$ we write simply $\Lambda$, which is representing a constant contribution to the $\Lambda(H)$ given by the expansion in the Taylor series (\ref{series}).

Now, with the help of the first integral (\ref{integral2}) we rewrite system (\ref{dyn4p})-(\ref{dyn5p}) to the new form
\begin{align}
x' &= 2\left(-x-\frac{1}{6}y+\frac{3x-y}{3}\right)=-y,\\
y' &= -3y-(\alpha_2+2\alpha_4 x+\cdots)\frac{3x-y}{9}.
\end{align}
Therefore all trajectories of the system on the plane ($x, y$) are determined by the first integral (\ref{integral2}).

The dynamical system (\ref{dyn4p})-(\ref{dyn5p}) at a finite domain has a critical point of the one type: a stationary solution $x=x_0$, $y=y_0=0$ representing a de Sitter universe. It can be stable and unstable and both lies on $x$ axis. Note that if stationary solutions exist than they always lie on the intersection $x$ axis ($y=0$) with the trajectory of the flat model represented by the first integral (\ref{integral2}), i.e. they are solutions of the following polynomial equation
\begin{equation}
x-\frac{1}{3}(\Lambda+\alpha_2 x+\alpha_4 x^2+\cdots)=0 \quad \text{and} \quad y=0 \quad \text{(empty universe).}
\end{equation}
Note that the static critical point which is representing the static Einstein universe does not satisfy the first integral (\ref{integral2}) because both $y$ and $\Lambda$ are positive.
Note also that if we substitute $y$ to (\ref{dyn4p}) then dynamics is reduced to the form of a one-dimensional dynamical system
\begin{align}
\frac{dx}{d\tau} &=-(3x-\Lambda-\alpha_2 x-\alpha_4 x^2 - \cdots). \label{dyn6p}\\
y &= 3x-(\Lambda+\alpha_2 x+\alpha_4 x^2+\cdots).
\end{align}

Following the Hartmann-Grobman theorem \cite{Perko:2001de} a system in neighborhood of critical points is well approximated by its linear part obtained by its linearization around of this critical point.

On the other hand because a linear part dominates for small $x$ in a right-hand side. Let us consider dynamic system (\ref{dyn6p}) truncated on this linear contribution then the Hartman-Grobman theorem \cite{Perko:2001de} guarantees us that dynamical system in vicinity of the critical point is a good approximation of the behaviour near the critical points. This system has the simple form
\begin{align}
\frac{dx}{d\tau} &=x(\alpha_2-3)+\Lambda,\label{dyn7} \\
y &=(3-\alpha_2)x-\Lambda. \label{dyn8}
\end{align}
The system (\ref{dyn7})-(\ref{dyn8}) has the single critical point of the form
\begin{equation}
x_0=\frac{\Lambda}{3-\alpha_2}, \quad y=0.
\end{equation}
It is representing an empty de Sitter universe.

\begin{figure}
	\centering
	\includegraphics[width=0.7\linewidth]{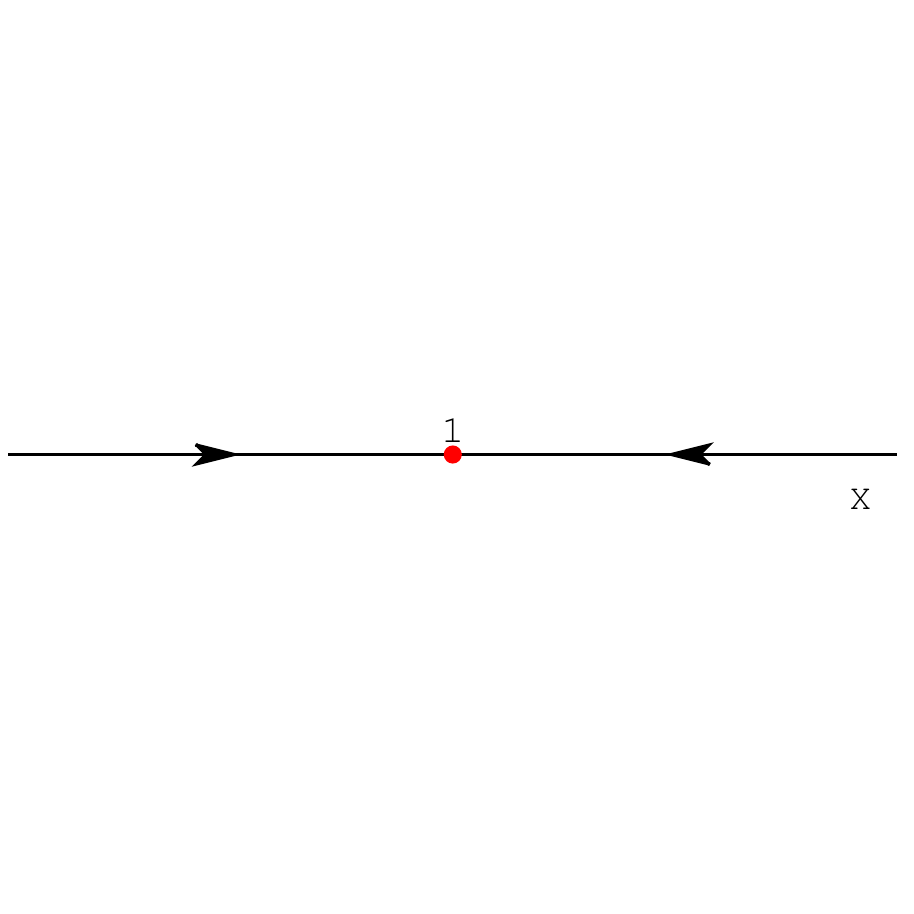}
	\caption{A one-dimensional phase portrait of the FRW model with $\Lambda=\Lambda(H)$. Note the existence of universal behaviour of the $H^2 (a)$ relation near the stable critical point (1) of the type of stable node. In neighborhood of this attractor we have the solution $X=H^2\pm\frac{\Lambda}{\alpha_2 +3}=X_0 a^{\alpha_2 -3}$ and $\rho_{\text{m}}=(3-\alpha_2)H^2-\Lambda=X_0 a^{\alpha_2-3}$. Therefore both $\rho_\text{m}$ and $\rho_\Lambda-\Lambda$ are proportional (scaling solution).}
	\label{fig:4}
\end{figure}

Let us now shift the position to this critical point to the origin, which one can perform after introducing the new variable $x\rightarrow X=x-x_0$. Then we obtain
\begin{equation}
\frac{dX}{d\tau}=(\alpha_2-3)X,
\end{equation}
which possesses the exact solution is of the form
\begin{equation}
X=X_0 e^{\tau(\alpha_2-3)}=X_0 a^{-3+\alpha_2},\label{sol}
\end{equation}
where $\alpha_2$ is constant. Of course this critical point is asymptotically stable if $\alpha_2<3$. The trajectories approaching this critical point at $\tau=\ln a\rightarrow\infty$ has attractor solution $X=X_0 a^{\alpha_2-3}$ or $x=X+x_0$, where $x_0=\frac{\Lambda}{3-\alpha_2}$ or $X=0$ (see Fig.~\ref{fig:4}). This attractor solution is crucial for the construction of a new model of decaying Lambda effect strictly connected with the dark matter problem \cite{Alcaniz:2005dg,Wang:2004cp}.

The solution (\ref{sol}) has natural interpretation: in a neighborhood of a global attractor of system (\ref{dyn6p}) trajectories behaves universal solution which motivates the Alcaniz-Lima approach in which
\begin{equation}
x=H^2=\frac{\tilde{\rho}_{\text{m},0}}{3}a^{-3+\alpha_2}+\frac{\rho_{\Lambda,0}}{3},
\end{equation}
where $\tilde{\rho}_{\text{m},0}=\frac{3}{3-\alpha_2}\rho_{\text{m},0}$.

\section{$\Lambda(a(t))$CDM cosmologies as a dynamical systems}
Many cosmological models of decaying $\Lambda$ in his construction makes the ansatz that $\Lambda(t)=\Lambda(a(t))$. For review of different approaches in which ansatzes of this type appeared see Table~\ref{tab:1}.

In this section, we would like to discuss some general properties of the corresponding dynamical systems modeling decaying $\Lambda$ term. It would be convenient to introduce the dynamical system in the state variables ($H,\rho$).

It has the following form
\begin{align}
\dot{H} &= -H^2-\frac{1}{6}\rho_{\text{m}}+\frac{\Lambda_{\text{bare}}}{3}+\frac{\Lambda(a)}{3},\label{1eq} \\
\dot\rho_{\text{m}} &= -3H\rho_{\text{m}}-\frac{d\Lambda}{da}(Ha)\label{eqadd}
\end{align}
or
\begin{align}
\frac{dH^2}{d \ln a} &= 2(-H^2-\frac{1}{6}\rho_{\text{m}}+\frac{1}{3}\Lambda_{\text{bare}}+\frac{1}{3}\Lambda(a)),\label{2eq}\\
\frac{d\rho_{\text{m}}}{d\tau} &= \frac{d\rho_{\text{m}}}{d\ln a} = -3\rho_{\text{m}}-a\frac{d\Lambda}{da}\label{7eq}
\end{align}
with first integral in the form
\begin{equation}
\rho_{\text{m}}=3H^2-\Lambda_{\text{bare}}-\Lambda(a).\label{integral3}
\end{equation}

\begin{table}[t]
	\caption{Different choices of $\Lambda(a)$ parametrization for different cosmological models appeared in literature.}
	\label{tab:1}
	\centering
	\begin{tabular}{ll}
		\hline
		$\Lambda(a)$ parametrization & References\\ \hline
		$\Lambda\sim a^{-m}$ & \cite{Jafarizadeh:1998wq,Silveira:1997fp}\\
		$\Lambda=M_{\text{pl}}^4 (\frac{r_{\text{pl}}}{R})^n$ & \cite{Chen:1990jw}\\
		$\Lambda=\frac{c^5}{\hbar G^2}(\frac{l_{pl}}{a})^n$ & \cite{John:1999gm}\\
		$\rho_\Lambda=\tilde\rho_{v,0}+\frac{\epsilon\rho_{\text{m},0}}{3-\epsilon}a^{-3+\epsilon}$ & \cite{Alcaniz:2005dg}\\
		$\rho_{de}=a^{-(4+\frac{2}{\alpha})}, \quad  \alpha=2(1-\Omega_{\text{\text{m}},0}-\Omega_{\Lambda,0})$& \cite{Gao:2007ep} \\
		$\rho_{\Lambda}=3\alpha^2M_p^2 a^{-2(1+\frac{1}{c})}$ & \cite{Li:2004rb}\\
		$\Lambda=\frac{\Lambda_{\text{pl}}}{(R/l_{\text{pl}})^2}\propto R^{-2}$ & \cite{Lopez:1995eb}\\ \hline
	\end{tabular}

\end{table}

\begin{figure}
	\centering
	\includegraphics[width=0.7\linewidth]{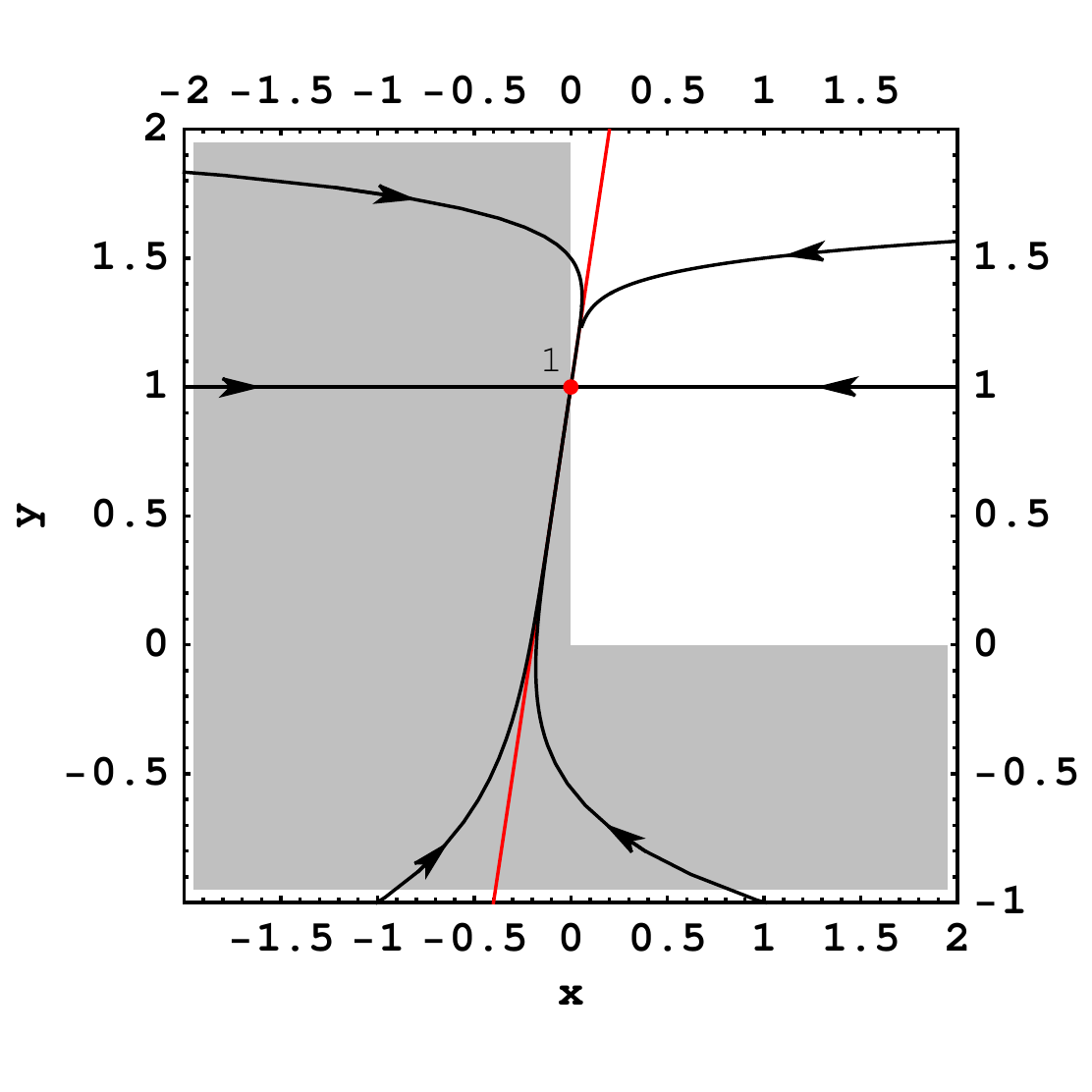}
	\caption{A phase portrait for dynamical system (\ref{dyn31x})-(\ref{dyn33x}). Critical point (1) is at $x=0$, $y=\Lambda$ is stable node. It is representing de Sitter universe. The red line represents the solution of scaling type $y=(3(3-\epsilon)-1)x+\Lambda$. Grey region represents non-physical domain excluded by the condition $\rho_{\text{m}}=x>0$, $\rho_{\Lambda}=y>0$. Note that trajectories approaches the attractor along the straight line. Let us note the existence of trajectories coming to the physical region from the nonphysical one. We treated this type of behaviour as a pathology related with appearance of ghost trajectories, which is emerging from the non-physical region.}
	\label{fig:2}
\end{figure}

\begin{figure}
	\centering
	\includegraphics[width=0.7\linewidth]{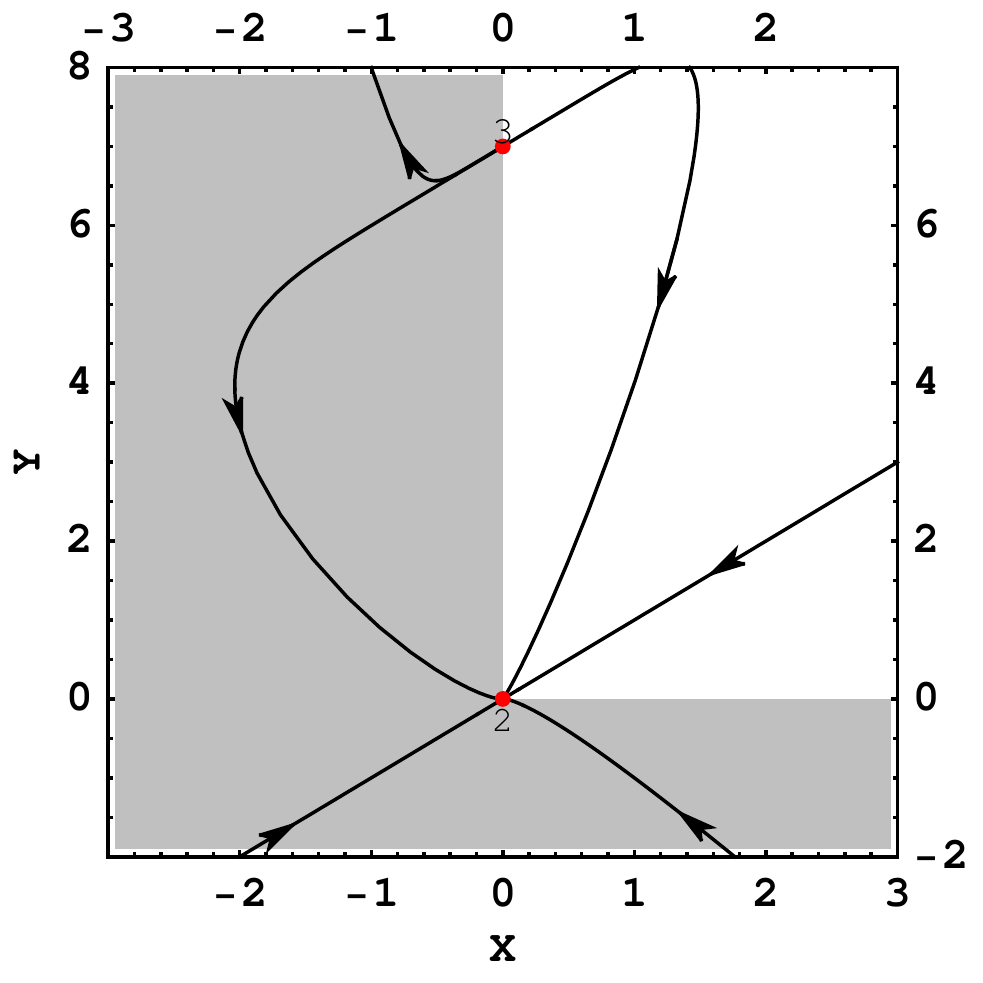}
	\caption{A phase portrait for dynamical system (\ref{dyn31})-(\ref{dyn33}). Critical point (2) at the origin $X=0$, $Y=0$ and critical point (3) is at $X=0$, $Y=10-3\epsilon$ present node. Grey region represents non-physical domain excluded by the condition $X>0$, $Y>0$.}
	\label{fig:5}
\end{figure}

\begin{figure}
	\centering
	\includegraphics[width=0.7\linewidth]{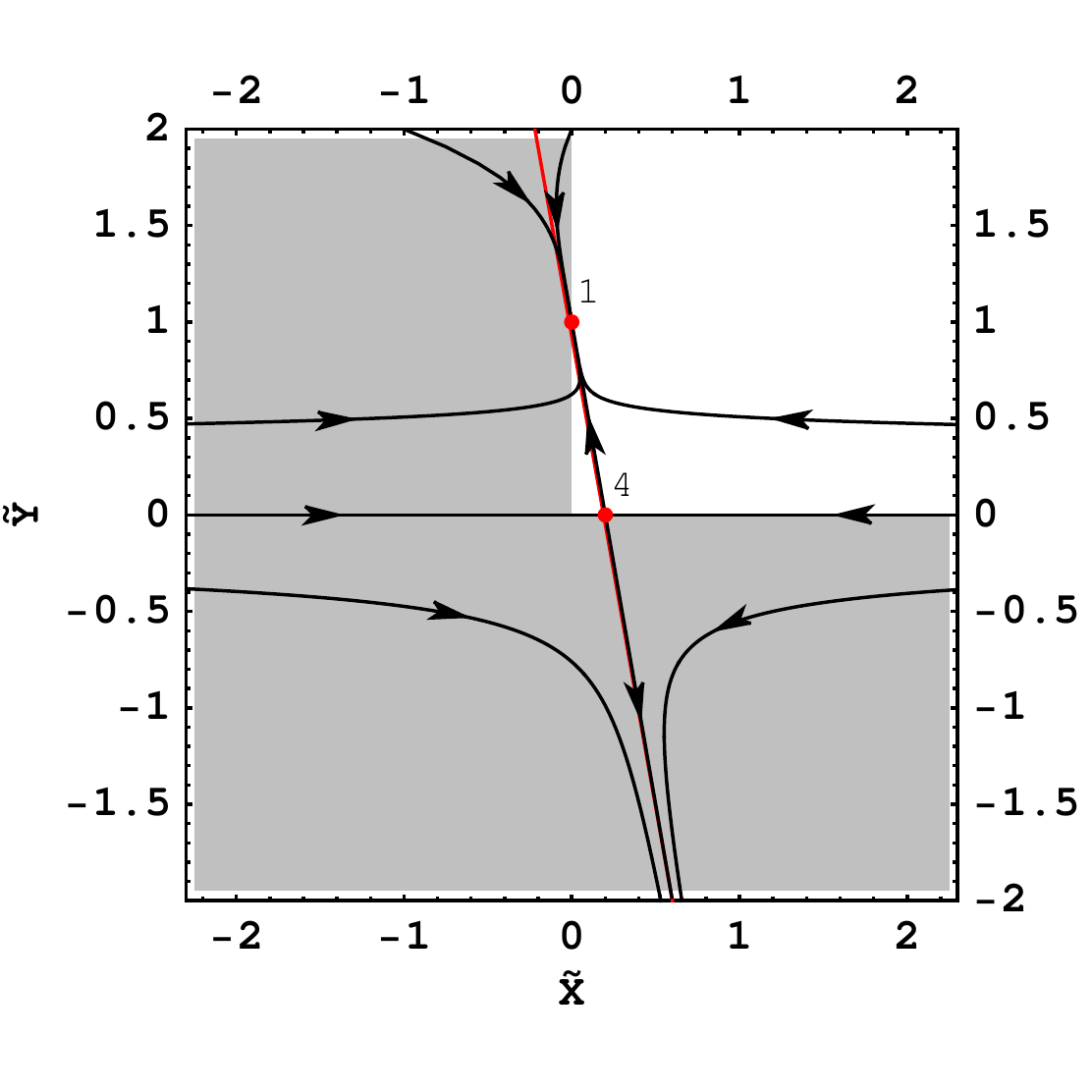}
	\caption{A phase portrait for dynamical system (\ref{dyn31y})-(\ref{dyn33y}). Critical point (1) is at $x=0$, $y=\Lambda$ presents stable node and critical point (4) is at  $\tilde{X}=\frac{1}{8-3\epsilon}$, $\tilde{Y}=0$ presents saddle type point. The red line represents solution $\tilde{Y}={3(3-\epsilon)-1+\Lambda\tilde{X}}$. Grey region represents non-physical domain excluded by the condition $\tilde{X}\tilde{Y}>0$.}
	\label{fig:6}
\end{figure}

\begin{figure}
	\centering
	\includegraphics[width=0.7\linewidth]{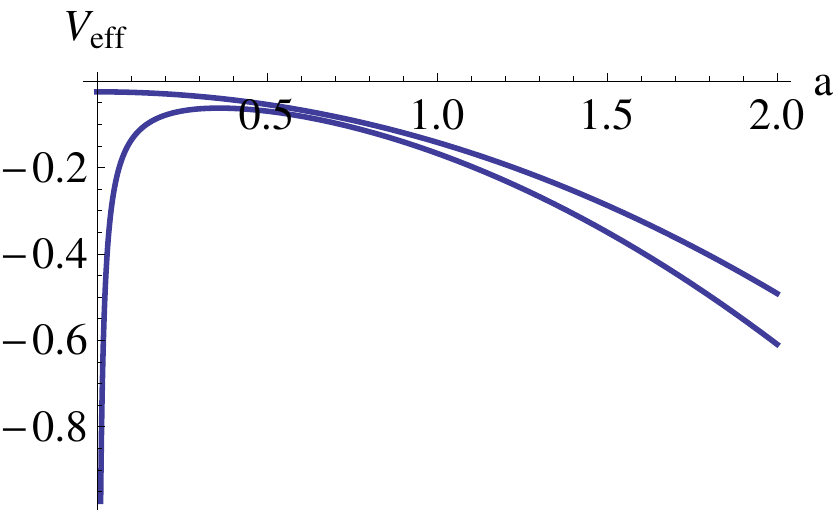}
	\caption{The figure presents a potential $V_{\text{eff}}(a)$. Top trajectory represents $V_{\text{eff}}$ for example $\epsilon=1$. Bottom trajectory represents $V_{\text{eff}}$ for example $\epsilon=0.1$. The shape of diagram of the potential determine the phase space structure. The maximum of the potential is corresponding static Einstein universe in the phase space.}
	\label{fig:3}
\end{figure}

If we have prescribed the form of $\Lambda(a)$ relation, then in the dynamical analysis we can start with the first equation (\ref{1eq}) which it would be convenient to rewrite to the form of acceleration equation, i.e.,
\begin{equation}
\frac{\ddot a}{a}=-\frac{1}{6}\rho_{\text{m}}(a)+\frac{\Lambda_{\text{bare}}}{3}+\frac{\Lambda(a)}{3},\label{3eq}
\end{equation}
where $\rho_{\text{m}}(a)$ is determined by equation (\ref{7eq}) which is a linear non-homogeneous differential equation which can solved analytically
\begin{equation}
\frac{d\rho_{\text{m}}}{d\tau}=-3\rho_{\text{m}}-\frac{d\Lambda}{d\tau}(a)
\end{equation}
and
\begin{equation}
\rho_{\text{m}}=-\left(\int^a a^3 d\Lambda(a)+C\right)a^{-3}
\end{equation}
Equation (\ref{3eq}) can be rewritten to the form analogous to the Newtonian equation of motion for particle of unit mass moving in the potential $V(a)$, namely
\begin{equation}
\ddot{a}=-\frac{\partial V(a)}{\partial a},\label{dda}
\end{equation}
where
\begin{equation}
V(a)=\frac{1}{6}a^{-2}\left(\int^a a^3\frac{d\Lambda}{da}da+C\right)+\frac{\Lambda_{\text{bare}}}{3}a+\frac{1}{3}a\Lambda(a).
\end{equation}
Integration of the above function gives form of the potential.

Of course, equation (\ref{dda}) can be rewritten to the Newtonian two-dimensional dynamical system
\begin{equation}
\dot a=p, \qquad \dot p=-\frac{\partial V}{\partial a},
\end{equation}
where
\begin{equation}
\frac{\dot a^2}{2}+V(a)=E=\text{const}.\label{energy}
\end{equation}
The integral of energy (\ref{energy}) should be consistent with the first integral (\ref{integral3}), i.e.
\begin{equation}
\rho_{\text{m}}+\rho_\Lambda=3H^2,
\end{equation}
\begin{equation}
a^{-3}\int a^3 \Lambda'(a)da+3\frac{\dot a^2}{a^2}=\Lambda_{\text{bare}}+\Lambda(a).
\end{equation}

Because the system under consideration is a conservative system, centres or saddles can appear in the phase. If the potential function possesses a maximum, then in the phase space we obtain a saddle type critical point and if $V(a)$ has a minimum this point is corresponding to a centre.

As an example of adopting the method of the effective potential, which is presented here, let us consider the parametrization of $\Lambda(a)$ like in the Alcaniz-Lima model of decaying vacuum \cite{Alcaniz:2005dg}. They assumed that energy density of vacuum is of the form (see Table~\ref{tab:1})
\begin{equation}
\rho_\Lambda=\rho_{v,0}+\frac{\epsilon\rho_{\text{m},0}}{3-\epsilon}a^{-3+\epsilon},\label{par}
\end{equation}
where $\rho_{v,0}$ is vacuum energy $\rho_{\text{m},0}$ is the energy density of matter at the present moment for which we choose $a=1=a_0$. Because $\dot{\rho}_{\text{vac}}<0$, i.e. the energy of vacuum is decaying from conservation conditions
\begin{equation}
\dot\rho_m=-3H\rho_m-\dot\rho_{\text{vac}}\label{vac}
\end{equation}
we obtain that
\begin{equation}
\dot\rho_{\text{vac}}=-\dot\rho_{\text{m}}-3H\rho_{\text{m}}=-\rho_{\text{m}}\left(\frac{\dot{\rho}_{\text{m}}}{\rho_{\text{m}}}+3H\right)
\end{equation}
and vacuum is decaying if
\begin{equation}
\frac{d\ln \rho_{\text{m}}}{d\ln a}>-3.
\end{equation}
Let us note that $\rho_{\text{m}}=0$ is a solution of system (\ref{vac}) only if $\Lambda$ is constant. It is a source of different pathologies in the phase space because the trajectories can pass through the line $\rho_{\text{m}}=0$. As a consequence of decaying vacuum energy density of matter will dilute more slowly compared to the corresponding canonical relation in the $\Lambda$CDM model, i.e. the energy density of matter is scaling following the rule
\begin{equation}
\rho_{\text{m}}=\rho_{\text{m},0}a^{-3+\epsilon},
\end{equation}
where $\epsilon>0$.

Dynamical system obtained from equation (\ref{2eq})-(\ref{7eq}) with parametrization (\ref{par}) has the following form
\begin{align}
x'&=-3x+\frac{y-\Lambda}{3-\epsilon},\label{dyn31}\\
y'&=-\frac{y-\Lambda}{3-\epsilon},\label{dyn33}\\
z'&=-z-\frac{x}{6}+\frac{y}{3},\label{dyn32}
\end{align}
with condition $y=\Lambda+\frac{\epsilon}{3-\epsilon}x$,
where $x=\rho_{\text{m}}$, $y=\rho_\Lambda$, $z=H^2$ and $'\equiv \frac{d}{d\tau}$. The above dynamical system contains the autonomous two-dimensional dynamical system (\ref{dyn31})-(\ref{dyn33}). Therefore this system has invariant two-dimensional submanifold. A phase portrait with this invariant submanifold is demonstrated in Fig.~\ref{fig:2}.

For deeper analysis of the system, the investigation of trajectories at the circle $x^2+y^2=\infty$ at infinity is required. For this aim dynamical system (\ref{dyn31})-(\ref{dyn33}) is rewritten in projective coordinates. Two maps ($X,Y$) and ($\tilde{X},\tilde{Y}$) cover the circle at infinity. In the first map we use following projective coordinates: $X=\frac{1}{x}$, $Y=\frac{y}{x}$ and in the second one $\tilde{X}=\frac{x}{y}$, $\tilde{Y}=\frac{1}{y}$. The system (\ref{dyn31})-(\ref{dyn33}) rewritten in coordinates $X$ and $Y$ has the following form
\begin{align}
X'&=X\left(\frac{Y-\Lambda X}{3-\epsilon}-3\right),\label{dyn31x}\\
Y'&=(Y-1)\frac{Y-\Lambda X}{3-\epsilon}-3Y\label{dyn33x}
\end{align}
and for variables $\tilde{X}$, $\tilde{Y}$, we obtain
\begin{align}
\tilde{X}'&=(1+\tilde{X})\frac{1-\Lambda \tilde{Y}}{3-\epsilon}-3\tilde{X},\label{dyn31y}\\
\tilde{Y}'&=\tilde{Y}\frac{1-\Lambda \tilde{Y}}{3-\epsilon}.\label{dyn33y}
\end{align}
The phase portraits for dynamical systems (\ref{dyn31x})-(\ref{dyn33x}) and (\ref{dyn31y})-(\ref{dyn33y}) are demonstrated in Figures \ref{fig:4} and \ref{fig:5}. Critical points for the above dynamical system are presented in Table \ref{tab:2}.

\begin{table}[t]
	\caption{Critical points for autonomous dynamical systems (\ref{dyn31})-(\ref{dyn33}), (\ref{dyn31x})-(\ref{dyn33x}), (\ref{dyn31y})-(\ref{dyn33y}), their eigenvalues and cosmological interpretation.}
	\label{tab:2}
	\centering
	\begin{tabular}{lllll}
		\hline
		No & critical point & eigenvalues & type of critical point & type of universe\\ \hline
		1 & $x=0$, $y=\Lambda$ & -3, $\frac{1}{\epsilon-3}$ & stable node & de Sitter\\
		2 & $X=0$, $Y=0$ & -3, $-3+\frac{1}{\epsilon-3}$ & stable node & Einstein-de Sitter\\
		3 & $X=0$, $Y=10-3\epsilon$ & $\frac{1}{3-\epsilon}$, $3+\frac{1}{3-\epsilon}$ & stable node & scaling universe\\
		&  &  &  & $\rho_{\text{m}}$ is proportional to $\rho_{\Lambda}$\\
		4 & $\tilde{X}=\frac{1}{8-3\epsilon}$, $\tilde{Y}=0$ & $\frac{1}{3-\epsilon}$, $-3+\frac{1}{3-\epsilon}$ & saddle & scaling universe\\ 
		&  &  &  & $\rho_{\text{m}}$ is proportional to $\rho_{\Lambda}$\\\hline
	\end{tabular}
\end{table}

The reduction of dynamics to the particle like description with the effective potential enable us to treat evolution of the universe like in classical mechanics in terms of the scale factor as a positional variable and
\begin{equation}
V_{\text{eff}}(a)=-\frac{\rho_{\text{eff}}(a)a^2}{6}=-\frac{1}{6}a^2\left(\rho_m a^{-3+\epsilon}+\rho_{v,0}+\frac{\epsilon\rho_{m,0}a^{-3+\epsilon}}{3-\epsilon}\right),
\end{equation}
where $\rho_{\text{eff}}=\rho_{\text{m}}+\rho_{\text{vac}}(a)$ and
\begin{equation}
\frac{\dot a^2}{2}+V_{\text{eff}}=-\frac{k}{2}.
\end{equation}

The motion of particle mimicking the evolution of the universe is restricted to the zero energy level $E=0$ (because we considered a flat model). The evolutionary paths for the model can be directly determined from the diagram of the effective potential $V_{\text{eff}}(a)$ function it self.

Fig.~\ref{fig:3} demonstrates the diagram $V_{\text{eff}}(a)$ for values $\epsilon=0.1$ and $1$. In general, for the phase portrait in the plane ($a, \dot a$) the maximum of $V(a)$ is corresponding to the static Einstein universe. This critical point is situated on the $a$-axis and it is always of the saddle type. Of course it is only admissible for closed universe. In this case a minimum corresponding critical point is of the centre type.

The Alcaniz-Lima model behaves in the phase space ($a, \dot{a}$) like the $\Lambda$CDM one \cite{Alcaniz:2005dg}. Trajectories are starting from ($a, \dot a)=(0, \infty$) (corresponding with the big bang singularity) and coming toward to the static universe and than evolving to the infinity. Note that if $0<\epsilon<1$ then qualitatively dynamics is equivalent to the $\Lambda$CDM model.

Equation (\ref{eqadd}) can be rewrite as
\begin{equation}
\dot{\rho}_{\text{m}}=-3H\rho_{\text{m}}-3H\rho_{\text{m}}\delta(t)+\frac{d\Lambda}{da}Ha,
\end{equation}
where $\delta(t)=-\frac{1}{\rho_{\text{m}}}\frac{d\Lambda}{da}a$.
Therefore
\begin{equation}
\dot{\rho}_{\text{m}}=-3H\rho_{\text{m}}(1+\delta(t)),\label{last}
\end{equation}
where $-3H\rho_{\text{m}}\delta(t)=\frac{d\Lambda}{da}Ha$, i.e.,
$\delta(t)=-\frac{\frac{d\Lambda}{da}a}{\rho_{\text{m}}}\propto-\frac{\rho_{\Lambda}}{\rho_{\text{m}}}$.
If $\delta(t)$ s slowly changing function of time, i.e., $\delta(t)\simeq\delta$ then (\ref{last}) has the solution $\rho_{\text{m}}=\rho_{\text{m,0}}a^{-3+\delta}$.

\section{$\Lambda(R)$CDM cosmologies as a dynamical systems}
Recently the Ricci scalar dark energy idea has been considered in the context of the holographic principle \cite{Cai:2007us}. In this case dark energy can depend on time $t$ through the Ricci scalar $R(t)$, i.e., $\Lambda(t)=\Lambda(R(t))$. Such a choice does not violate covariantness of general relativity. The special case is parameterization $\rho_{\Lambda}=3\alpha R=18\alpha(\dot H+2H^2+\frac{k}{a^2})$ \cite{Gao:2007ep}. Then cosmological equations are also formulated in the form of a two-dimensional dynamical system
\begin{align}
\dot H &=-H^2-\frac{1}{6}(\rho_{\text{m}}+\rho_{\Lambda}),\\
\dot{\rho} &=-3H\rho_{\text{m}}
\end{align}
with the first integral in the form
\begin{equation}
H^2=\frac{1}{3}\left(-\frac{3k}{a^2}+\frac{2}{2-\alpha}\rho_{\text{m},0}a^{-3}+f_0 a^{2\frac{1-2\alpha}{\alpha}}\right),
\end{equation}
where $f_0$ is an integration constant.

\begin{figure}
	\centering
	\includegraphics[width=0.7\linewidth]{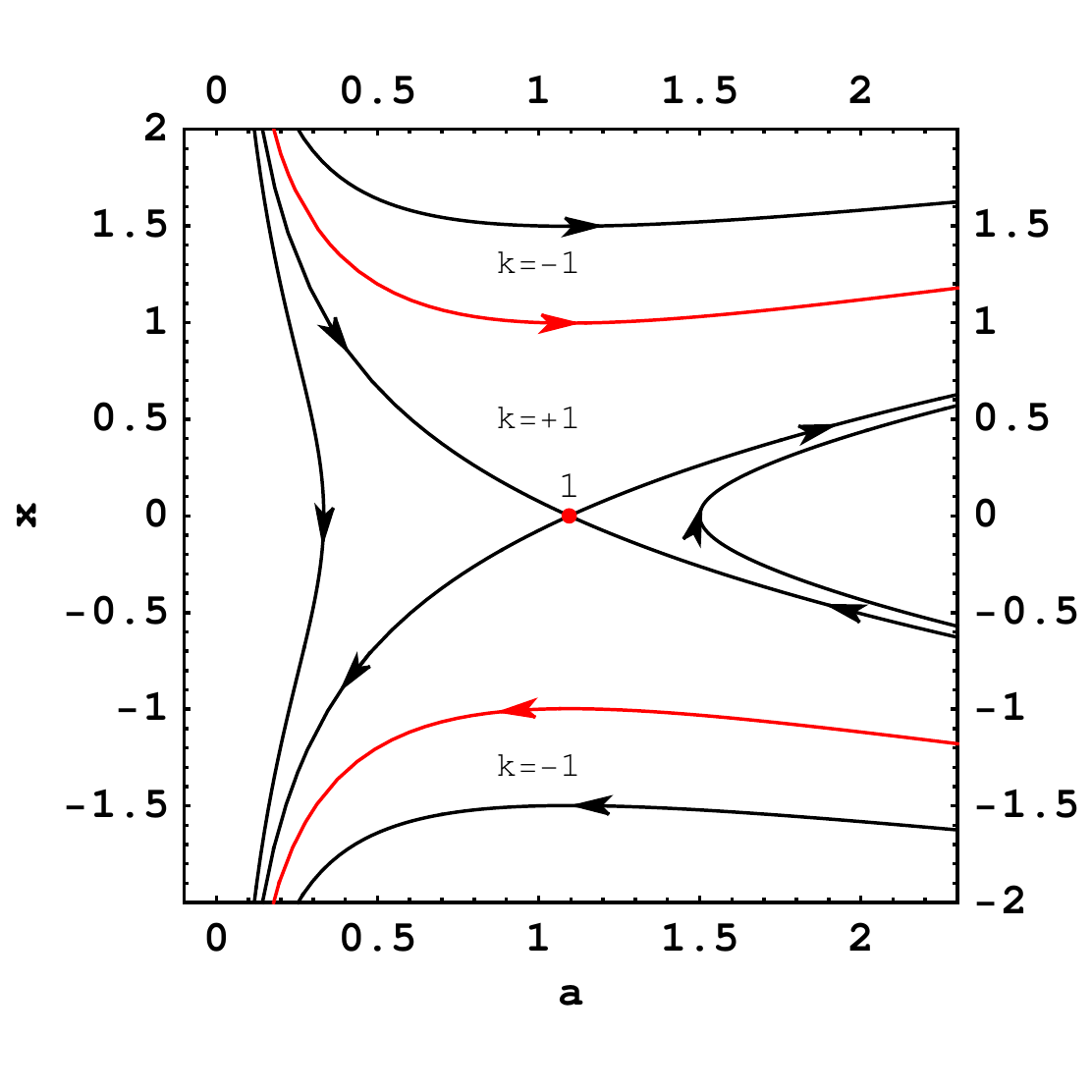}
	\caption{A phase portrait for dynamical system (\ref{riccia})-(\ref{riccix}) with $\alpha=2/3$. Critical point (1), which is located on $a$-axis $a=\sqrt{\frac{3\rho_{\text{m,0}}}{2\rho_{\Lambda}-\rho_{\text{m,0}}}}$, $x=0$, presents saddle point and which represents static Einstein universe. Red lines represent the trajectories of the flat universe. These trajectory separates the region in which lies closed and open models. Let us note topological equivalence of phase space trajectories of both presented in Fig.~\ref{fig:11} and Fig.~\ref{fig:9} if direction of time is omitted, i.e. $a\rightarrow\frac{1}{a}$ and $t\rightarrow -t$ is symmetry of the phase portrait. Note that when $\alpha=\frac{1}{2}$ then the phase portrait is equivalent $\Lambda$CDM one.}
	\label{fig:11}
\end{figure}

\begin{figure}
	\centering
	\includegraphics[width=0.7\linewidth]{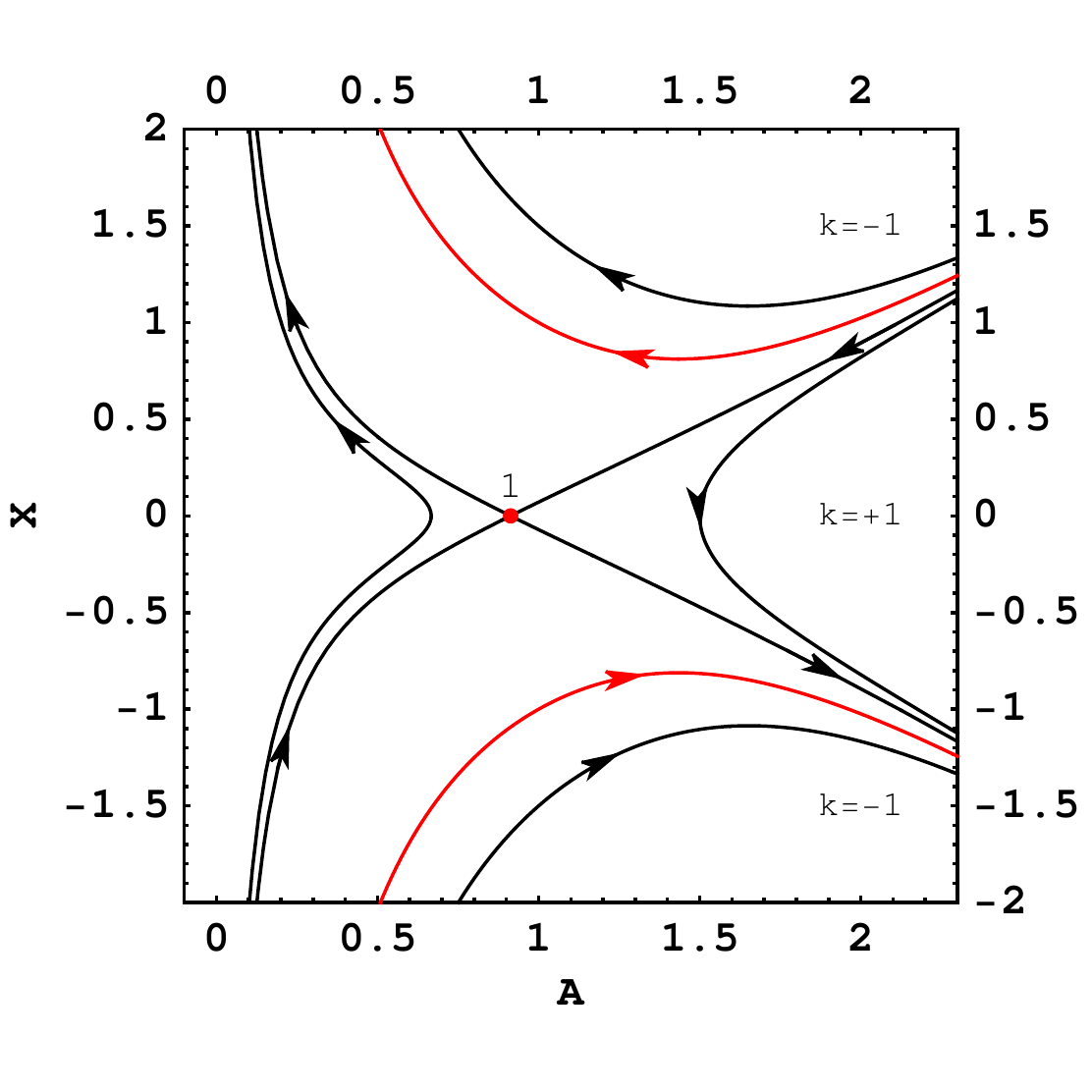}
	\caption{A phase portrait for dynamical system (\ref{ricciA})-(\ref{ricciX}) with $\alpha=2/3$. Critical point (1) on the $A$-axis $A=\sqrt{\frac{2\rho_{\Lambda}-\rho_{\text{m,0}}}{3\rho_{\text{m,0}}}}$, $X=0$ presents saddle and represents static Einstein universe. Red lines represent the trajectories of a flat universe and this trajectory separates the region in which lies closed and open models.}
	\label{fig:9}
\end{figure}

\begin{figure}
	\centering
	\includegraphics[width=0.7\linewidth]{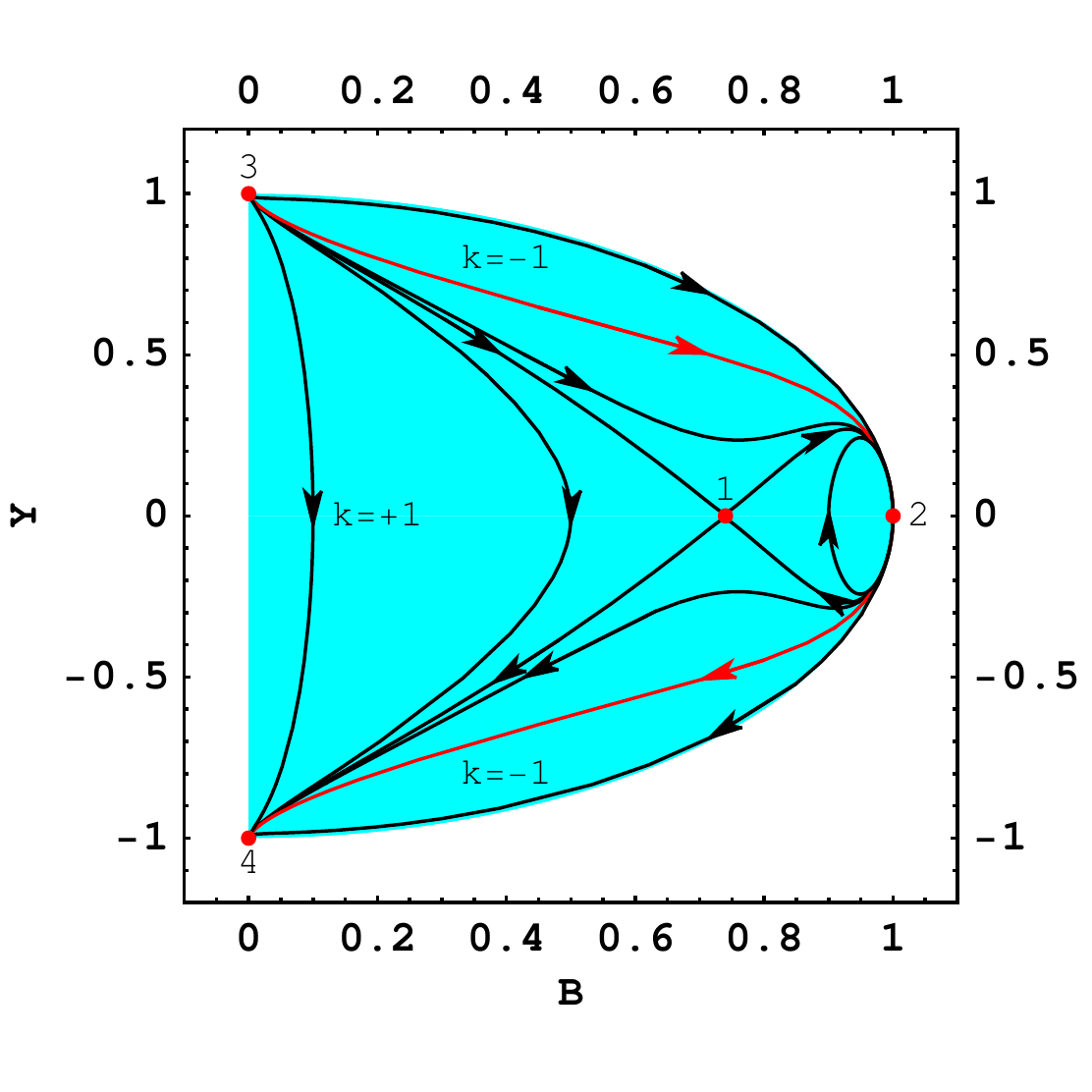}
	\caption{A phase portrait for dynamical system (\ref{ricciB})-(\ref{ricciY}) with $\alpha=2/3$. Critical point (1) is at $B=\sqrt{\frac{3\rho_{\text{m},0}}{2\rho_{\Lambda}-\rho_{\text{m},0}}+1}$, $Y=0$ presents saddle and represents static Einstein universe. Critical point (2) at the $B$-axis, $Y=0$ presents stable node and represents Milne universe. Critical points (3) and (4) at $B=0$, $Y=1$ and $B=0$, $Y=-1$  present nodes and represent Einstein-de Sitter universe. Blue region represents physical domain restricted $B^2 + Y^2 \le 0$, $B\ge 0$. Red lines represent the flat universe and these trajectories separates the region in which lies closed and open models.}
	\label{fig:10}
\end{figure}

From above equations, we can obtain a dynamical system in the state variables $a$, $x=\dot{a}$
\begin{align}
\dot{a}&=x,\label{riccia}\\
\dot{x}&=-\Omega_{\text{m,0}}\frac{1}{2-\alpha}a^{-2}+\left(\frac{1}{\alpha}-1\right)\left(\Omega_{\Lambda,0}-\Omega_{\text{m},0}\frac{\alpha}{2-\alpha}\right)a^{\frac{2}{\alpha}-3}.\label{riccix}
\end{align}
The phase portrait on the plane $(a, x)$ are presented in Fig.~\ref{fig:11}.

For the analysis the behaviour of trajectories at infinity we use the following sets of projective coordinates: $A=\frac{1}{a}$, $X=\frac{x}{a}$.

The dynamical system for variables $A$ and $X$ is expressed by
\begin{align}
\dot A&=-XA,\label{ricciA}\\
\dot{X}&=A^{3}\left(-\Omega_{\text{m,0}}\frac{1}{2-\alpha}+\left(\frac{1-\alpha}{\alpha}\right)\left(\Omega_{\Lambda,0}-\Omega_{\text{m},0}\frac{\alpha}{2-\alpha}\right)A^{\frac{\alpha-2}{\alpha}}\right)+X^2.\label{ricciX}
\end{align}
We can use also the Poincar{\'e} sphere to search critical points in the infinity. We introduce the following new variables: $B=\frac{a}{\sqrt{1+a^2+x^2}}$, $Y=\frac{x}{\sqrt{1+a^2+x^2}}$.
In variables $B$, $Y$, we obtain dynamical system in the form
\begin{equation}
\begin{array}{l}
B'=YB^2 (1-B^2)-\\
BY\left(-\Omega_{\text{m,0}}\frac{1}{2-\alpha}(1-B^2-Y^2)^{3/2}+\left(\frac{1-\alpha}{\alpha}\right)\left(\Omega_{\Lambda,0}-\Omega_{\text{m},0}\frac{\alpha}{2-\alpha}\right)B^{-1+2/\alpha}(1-B^2-Y^2)^{2-1/\alpha}\right),
\end{array}
\label{ricciB}
\end{equation}
\begin{equation}
\begin{array}{l}
Y'=\\ \left(-\Omega_{\text{m,0}}\frac{1}{2-\alpha}(1-B^2-Y^2)^{3/2}+\left(\frac{1-\alpha}{\alpha}\right)\left(\Omega_{\Lambda,0}-\Omega_{\text{m},0}\frac{\alpha}{2-\alpha}\right)B^{-1+2/\alpha}(1-B^2-Y^2)^{2-1/\alpha}\right)(1-Y^2)\\-Y^2 B^3,
\end{array}\label{ricciY}
\end{equation}
where $'\equiv B^2\frac{d}{dt}$.

A phase portraits for above dynamical systems are demonstrated in Figures \ref{fig:9} and \ref{fig:10}.

\section{Cosmology with emergent $\Lambda(a)$ relation from exact dynamics}
For illustration of an idea of emergent $\Lambda(a)$ relation let us consider cosmology with a scalar field which is non-minimal coupled to gravity. For simplicity without loss of generality of our consideration we assume that the non-minimal coupling $\xi$ is constant like the conformal coupling. It is also assumed dust matter, present in the model, does not interact with the scalar field. Because we would like to nest the $\Lambda$CDM model in our model we postulate that the potential of the scalar field is constant. We also assume flat geometry with the R-W metric. The action for our model assumes the following form
\begin{equation}
S=S_{\text{g}}+S_{\phi}+S_{\text{m}},
\end{equation}
where
\begin{align}
S_{\text{g}}+S_{\phi}&=\frac{1}{2}\int \sqrt{g}\left(R+g^{\mu\nu}\partial_{\mu}\phi\partial_{\nu}\phi-\xi R\phi^2-2V(\phi)\right)d^4 x, \\
S_{\text{m}}&= \int\sqrt{g}\mathcal{L}_{\text{m}}d^4 x,
\end{align}
where the metric signature is $(-,+,+,+)$, $R=6\left(\frac{\ddot a}{a}+\frac{\dot a^2}{a^2}\right)$ is the Ricci scalar and a dot denotes differentiation with respect the cosmological time $t$, i.e., $\dot{}\equiv\frac{d}{dt}$ and $\mathcal{L}_{\text{m}}=-\rho\left(1+\int\frac{p(\rho)}{\rho^2}d\rho\right)$.

After dropping the full derivatives with respect to the time the equation of motion for the scalar field are obtained after variation over scalar field and metric
\begin{equation}
\frac{\delta S}{\delta\phi}=0 \quad \Leftrightarrow \quad \ddot \phi+3H\dot{\phi}+\xi R\phi+V'(\phi)=0,
\end{equation}
where $'\equiv\frac{d}{d\phi}$ and
\begin{equation}
\frac{\delta S}{\delta g}=0 \quad \Leftrightarrow \quad  \mathcal{E}=\frac{1}{2}\dot\phi^2+3\xi H^2\phi^2+6\xi H\phi\dot{\phi}+V(\phi)-3H^2\equiv 0.\label{xi}
\end{equation}

Additionally from the conservation condition for the satisfying equation of state $p=p(\rho)$ of barotropic matter we have
\begin{equation}
\dot{\rho}=-3H(\rho+p(\rho)).\label{con}
\end{equation}
Because we assume dust matter ($p$=0) equation (\ref{con}) has a simple scaling solution of the form
\begin{equation}
\rho_{\text{m}}=\rho_{\text{m,0}}a^{-3},
\end{equation}
where $a=a(t)$ is the scale factor from the R-W metric $ds^2=dt^2-a^2(t)(dx^2+dy^2+dz^2)$.

Analogously the effects of the homogeneous scalar field satisfy the conservation condition
\begin{equation}
\dot{\rho}_{\phi}=-3H(\rho_{\phi}+p_{\phi}),
\end{equation}
where
\begin{align}
\rho_{\phi}&=\frac{1}{2}\dot\phi^2+V(\phi)+6\xi H\phi\dot{\phi}+3\xi H^2\phi^2,\\
p_{\phi}&=\frac{1}{2}(1-4\xi)\dot{\phi^2}-V(\phi)+2\xi H\phi\dot{\phi}-2\xi(1-6\xi)\dot H\phi^2-3\xi(1-8\xi)H^2\phi^2+2\xi\phi V'(\phi).
\end{align}

In the investigation of the dynamics it would be convenient to introduce so-called energetic state variables~\cite{Szydlowski:2008in}
\begin{equation}
x\equiv\frac{\dot{\phi}}{\sqrt{6}H}, \text{ } y\equiv\frac{\sqrt{V(\phi)}}{\sqrt{3}H}, \text{ } z\equiv\frac{\phi}{\sqrt{6}}.\label{var}
\end{equation}
The choice of such a form of state variables (\ref{var}) is suggested by the energy constraint $\mathcal{E}=0$ (\ref{xi}).

The energy constraint condition can be rewritten in terms of dimensionless density parameters
\begin{equation}
\Omega_{\text{m}}+\Omega_{\phi}=1\quad\Rightarrow\quad\Omega_{\phi}=1-\Omega_{\text{m}}=(1-6\xi)x^2+y^2+6\xi(x+z)^2=1-\Omega_{\text{m},0}a^{-3}
\end{equation}
and the formula $H(x,y,z,a)$ rewritten in terms of state variables $x$, $y$, $z$  assumes the following form
\begin{equation}
\left(\frac{H}{H_0}\right)^2=\Omega_{\phi}+\Omega_{\text{m}}=(1-6\xi)x^2+y^2+6\xi(x+z)^2+\Omega_{\text{m},0}a^{-3}.\label{friedmann}
\end{equation}
Formula (\ref{friedmann}) is crucial in the model testing and estimations of the model parameters by astronomical data.

Because we try to generalize the $\Lambda$CDM model it is natural to interpret the additional contribution beyond $\Lambda_{\text{bare}}$ as a running $\Lambda$ term in (\ref{friedmann}). In our further analysis we will called this term `emergent $\Lambda$ term'. Therefore
\begin{equation}
\Omega_{\Lambda,{\text{emergent}}}=(1-6\xi)x^2+y^2+6\xi(x+z)^2.\label{em}
\end{equation}
Of course state variables satisfy a set of the differential equations in the consequence of Einstein equations. We try to organize them in the form of autonomous differential equations, i.e., some dynamical system.

For this aim let us start from the acceleration equation
\begin{equation}
\dot H=-\frac{1}{2}(\rho_{\text{eff}}+p_{\text{eff}})=-\frac{3}{2}H^2(1+w_{\text{eff}}),
\end{equation}
where $\rho_{\text{eff}}$ and $p_{\text{eff}}$ are effective energy density and pressure and $w_{\text{eff}}=\frac{p_{\text{eff}}}{\rho_{\text{eff}}}$ is an effective coefficient of equation of state $\rho_{\text{eff}}=\rho_{\text{m}} +\rho_{\phi}$ and $p_{\text{eff}}=0+p_{\phi}$.

The coefficient equation of state $w_{\text{eff}}$ is given by formula
\begin{equation}
w_{\text{eff}}=(1-4\xi)x^2-y^2(1+2\xi\lambda z)+4\xi xz+12\xi^2 z^2,
\end{equation}
where $\lambda\equiv-\sqrt{6}\frac{V'(\phi)}{V(\phi)}$ is related with geometry of the potential, where $'\equiv\frac{d}{d\phi}$.

The dynamical system which is describing the evolution in phase space is in the form
\begin{align}
\frac{dx}{d(\ln a)}&=\frac{dx}{d\tau}=-3x-12\xi z+\frac{1}{2}\lambda y^2-(x+6\xi z)\frac{\dot H}{H^2},\label{dyn1}\\
\frac{dy}{d(\ln a)}&=\frac{dy}{d\tau}=-\frac{1}{2}\lambda xy-y\frac{\dot H}{H^2},\\
\frac{dz}{d(\ln a)}&=\frac{dz}{d\tau}=x,\\
\frac{d\lambda}{d(\ln a)} &=\frac{d\lambda}{d\tau}=-\lambda^2(\Gamma(\lambda)-1)x,\label{dyn2}
\end{align}
where $\Gamma=\frac{V''(\phi)V(\phi)}{V'^2(\phi)}$ and
\begin{multline}
\frac{\dot H}{H^2}=\frac{1}{H^2}\left[-\frac{1}{2}(\rho_{\phi}+p_{\phi})-\frac{1}{2}\rho_{\text{m,0}}a^{-3}\right]=\\
=\frac{1}{6\xi z^2 (1-6\xi)-1} \left[-12\xi(1-6\xi)z^2-3\xi\lambda y^2 z+\frac{3}{2}(1-6\xi)x^2+3\xi(x+z)^2+\frac{3}{2}-\frac{3}{2}y^2\right]. \label{H}
\end{multline}
Let us note that dynamical system (\ref{dyn1})-(\ref{dyn2}) is closed if we only assume that $\Gamma=\Gamma(\lambda)$.

From the form of the system (\ref{dyn1})-(\ref{dyn2}) one can observe that it has the invariant submanifold $\left\{\frac{\dot{H}}{H^2}=0\right\}$ for which the equation in the phase space is of the form
\begin{equation}
-12\xi(1-6\xi)z^2-3\xi\lambda y^2 z+\frac{3}{2}(1-6\xi)x^2+3\xi(x+z)^2+\frac{3}{2}-\frac{3}{2}y^2=0.
\end{equation}
Therefore there is no trajectories which intersect this invariant surface in the phase space. From the physical point of view the trajectories are stationary solutions and on this invariant submanifold they satisfy the condition
\begin{equation}
\frac{\dot H}{H^2}=0 \quad \Leftrightarrow \quad -\frac{1}{2}(\rho_{\phi}+p_{\phi})-\frac{1}{2}\rho_{\text{m},0}a^{-3}=0.
\end{equation}
If we look at the trajectories in the whole phase in neighborhood of this invariant submanifold, then we can observe that they will be asymptotically reached at an infinite value of time $\tau=\ln a$. They are tangent asymptotically to this surface. Note that in many cases the system on this invariant submanifolds can be solved and the exact solutions can be obtained.

As an illustration of the idea of the emergent $\Lambda(a)$ relation we consider two cases of cosmologies for which we derive $\Lambda=\Lambda(a)$ formulas. Such parametrizations of $\Lambda(a)$ arise if we consider a behaviour of trajectories near the invariant submanifold of dynamical systems
\begin{enumerate}
	\item $V=\text{const} \quad \text{or} \quad \lambda=0, \text{ case of } \xi=0$ \quad --- minimal coupling,
	\item $V=\text{const}, \quad \text{ case of } \xi=\frac{1}{6}$ \quad --- conformal coupling.
\end{enumerate}
In these cases the dynamical system (\ref{dyn1})-(\ref{dyn2}) reduces to
\begin{align}
	\frac{dx}{d(\ln a)}&=\frac{dx}{d\tau}=-3x-x\frac{\dot H}{H^2},\label{dyn3y}\\
	\frac{dy}{d(\ln a)}&=\frac{dy}{d\tau}=-y\frac{\dot H}{H^2},\label{dyn2y}\\
	\frac{dz}{d(\ln a)}&=\frac{dz}{d\tau}=x,\label{dyn4y}
\end{align}
where
\begin{align}
	\frac{\dot H}{H^2}=-\frac{3}{2}x^2-\frac{3}{2}+\frac{3}{2}y^2. \label{H2}
\end{align}
and
\begin{align}
\frac{dx}{d\tau}&=-3x-2z-\frac{\dot H}{H^2}(x+z),\label{dyn3}\\
\frac{dy}{d\tau}&=-y\frac{\dot H}{H^2},\label{dyn9}\\
\frac{dz}{d\tau}&=x,\label{dyn4}
\end{align}
where
\begin{equation}
\frac{\dot H}{H^2}=-\frac{1}{2}(x+z)^2-\frac{3}{2}+\frac{3}{2}y^2.
\end{equation}

Dynamical system (\ref{dyn3})-(\ref{dyn4}) can be rewrite in variable $X=x+z$, $Y=y$ and $Z=z$. Then we get
\begin{align}
	\frac{dX}{d\tau}&=-2X-\frac{\dot H}{H^2}X,\label{dyn3u}\\
	\frac{dY}{d\tau}&=-Y\frac{\dot H}{H^2},\label{dyn5u}\\
	\frac{dZ}{d\tau}&=X-Z,\label{dyn4u}
\end{align}
where
\begin{equation}
	\frac{\dot H}{H^2}=-\frac{1}{2}X^2-\frac{3}{2}+\frac{3}{2}Y^2.
\end{equation}

The next step in realization of our idea of the emergent $\Lambda$ is to solved the dynamical systems on an invariant submanifold and then to substitute this solution to formula (\ref{em}).

For the first case ($\xi=0, V=\text{const}$), dynamical system (\ref{dyn3y})-(\ref{dyn4y}) has the following form
\begin{align}
\frac{dx}{d(\ln a)}&=\frac{dx}{d\tau}=-3x,\label{dynx1}\\
\frac{dy}{d(\ln a)}&=\frac{dy}{d\tau}=0,\\
\frac{dz}{d(\ln a)}&=\frac{dz}{d\tau}=x,\label{dynx2}
\end{align}
with condition
\begin{align}
0=x^2-y^2+1.\label{con3}
\end{align}

\begin{figure}[ht]
	\centering
	\includegraphics[width=0.7\linewidth]{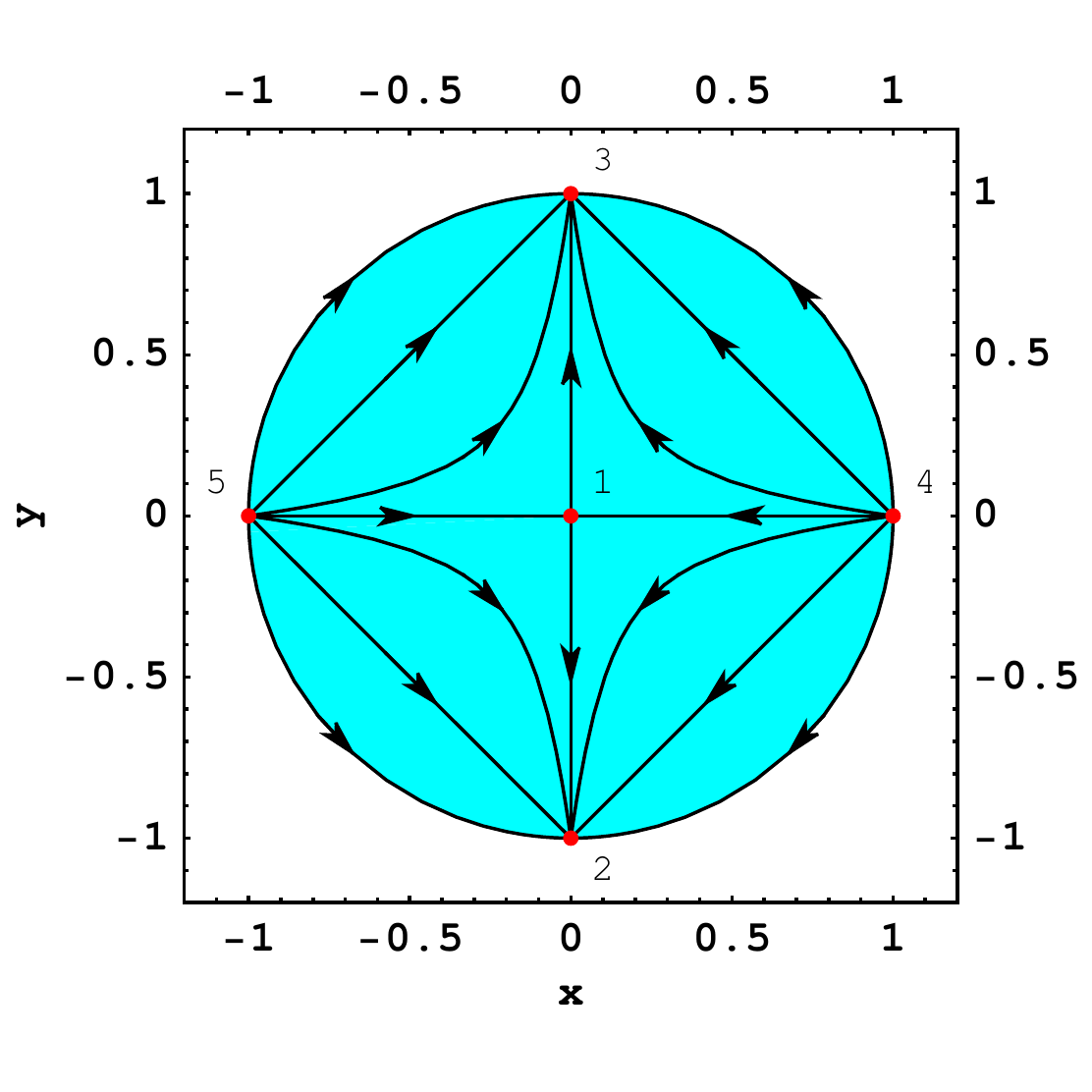}
	\caption{The phase portrait for autonomous dynamical system (\ref{dyn3y})-(\ref{dyn2y}). Critical points (1) represents a Einstein-de Sitter universe. Critical point (4) and (5) represent a Zeldovitch stiff matter universe. Critical points (2) represents a contracting  de Sitter. Critical points (3) represents stable de Sitter universe. The de Sitter universe is located on the invariant submanifold $\frac{\dot{H}}{H^2}=0$. The blue region presents physical region restricted by condition $x^2+y^2\le 1$, which is a consequence of $\Omega_{\text{m}} \ge 0$.}
	\label{fig:8}
	\end{figure}

\begin{figure}[ht]
	\centering
	\includegraphics[width=0.7\linewidth]{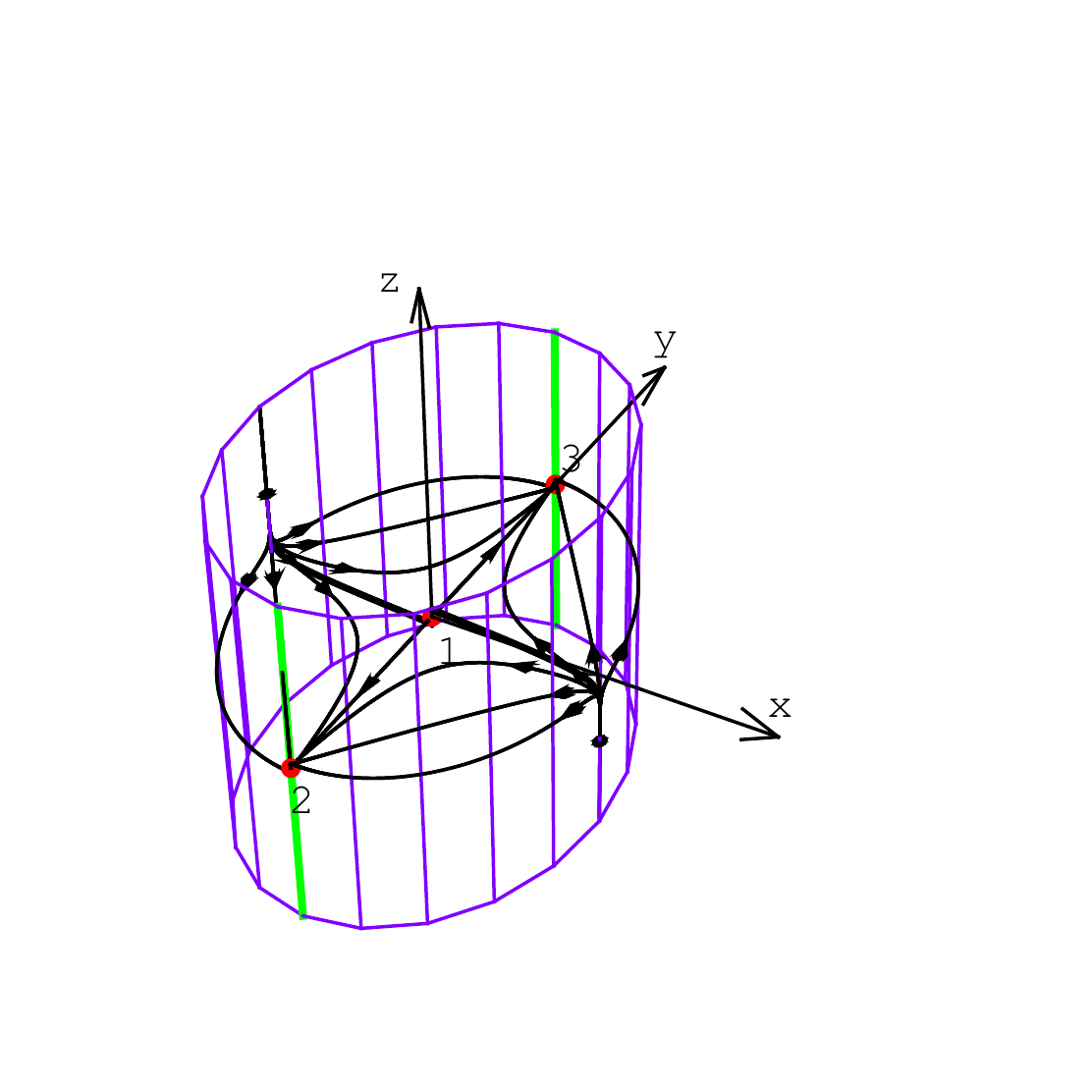}
	\caption{The phase portrait for dynamical system (\ref{dyn3y})-(\ref{dyn4y}). Critical points (1) represents the Einstein-de Sitter universe. Note that time $dt = Hd\tau$ is measured along trajectories, therefore in the region $H<0$ (contracting model) time $\tau$ is reversed to the original time $t$. Hence, critical point (2) represents an unstable de Sitter. Point (3) is opposite to critical points (2) which represents a contarcting de Sitter universe. The de Sitter universe is located on the invariant submanifold $\frac{\dot{H}}{H^2}=0$, which is the element of a cylinder and is presented by green lines. The surface of the cylinder presents a boundary of physical region restricted by condition $x^2+y^2\le 1$, which is a consequence of $\Omega_{\text{m}} \ge 0$.}
	\label{fig:18}
\end{figure}

\begin{figure}[ht]
	\centering
	\includegraphics[width=0.7\linewidth]{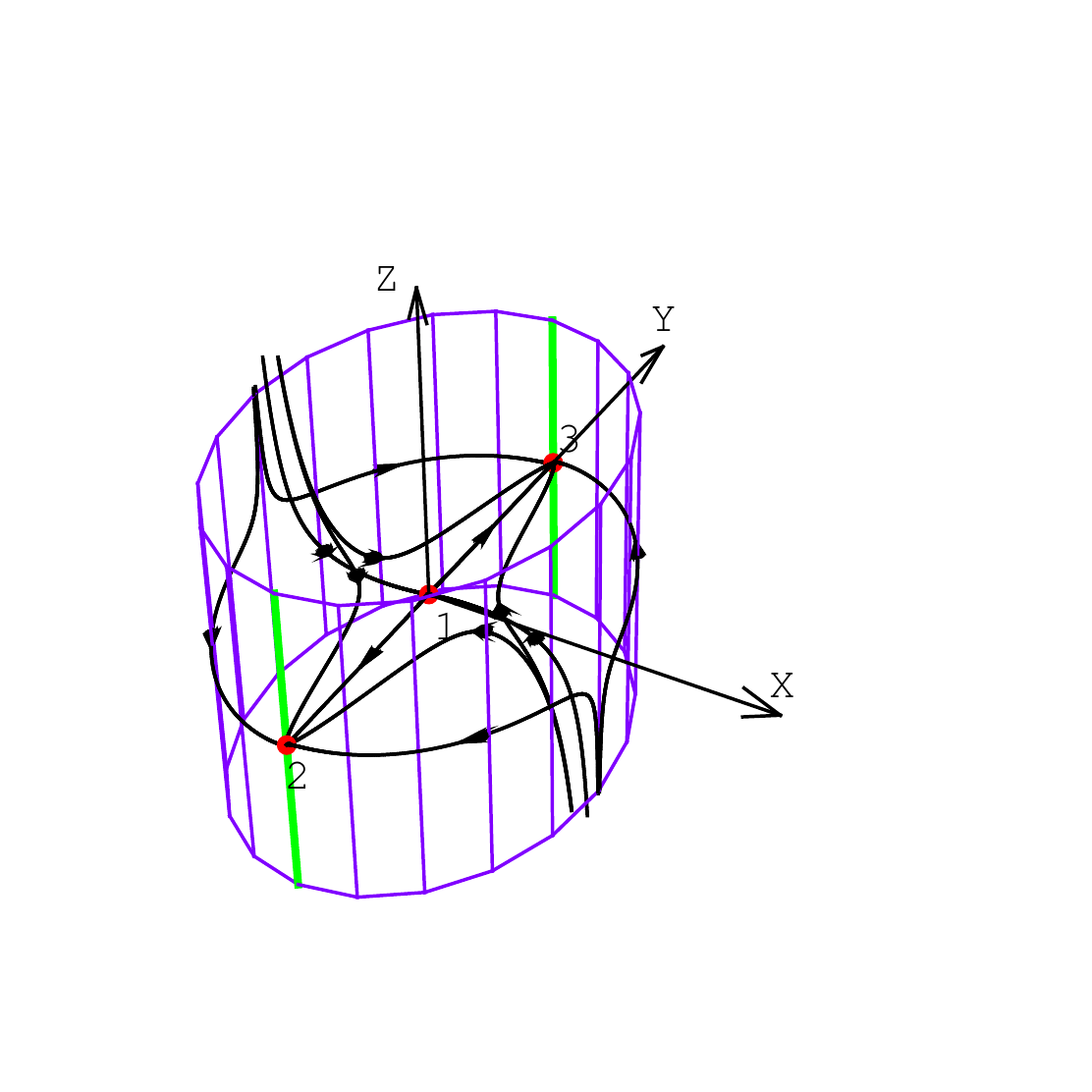}
	\caption{The phase portrait for dynamical system (\ref{dyn3u})-(\ref{dyn4u}). Critical point (1) represents an Einstein-de Sitter universe. Note that time $d\tau = Hdt $ is measured along trajectories, therefore in the region $H<0$ (contracting model) time $\tau$ is reversed to the original time $t$. Hence, critical point (2) represents an unstable de Sitter. Point (3) is opposite to critical points (2) which represents a contracting de Sitter universe. The de Sitter universe is located on the invariant submanifold $\frac{\dot{H}}{H^2}=0$ which is the element of a cylinder and is presented by green lines. The surface of the cylinder presents a boundary of physical region restricted by condition $X^2+Y^2\le 1$, which is a consequence of $\Omega_{\text{m}} \ge 0$.}
	\label{fig:19}
\end{figure}

\begin{figure}[ht]
	\centering
	\includegraphics[width=0.7\linewidth]{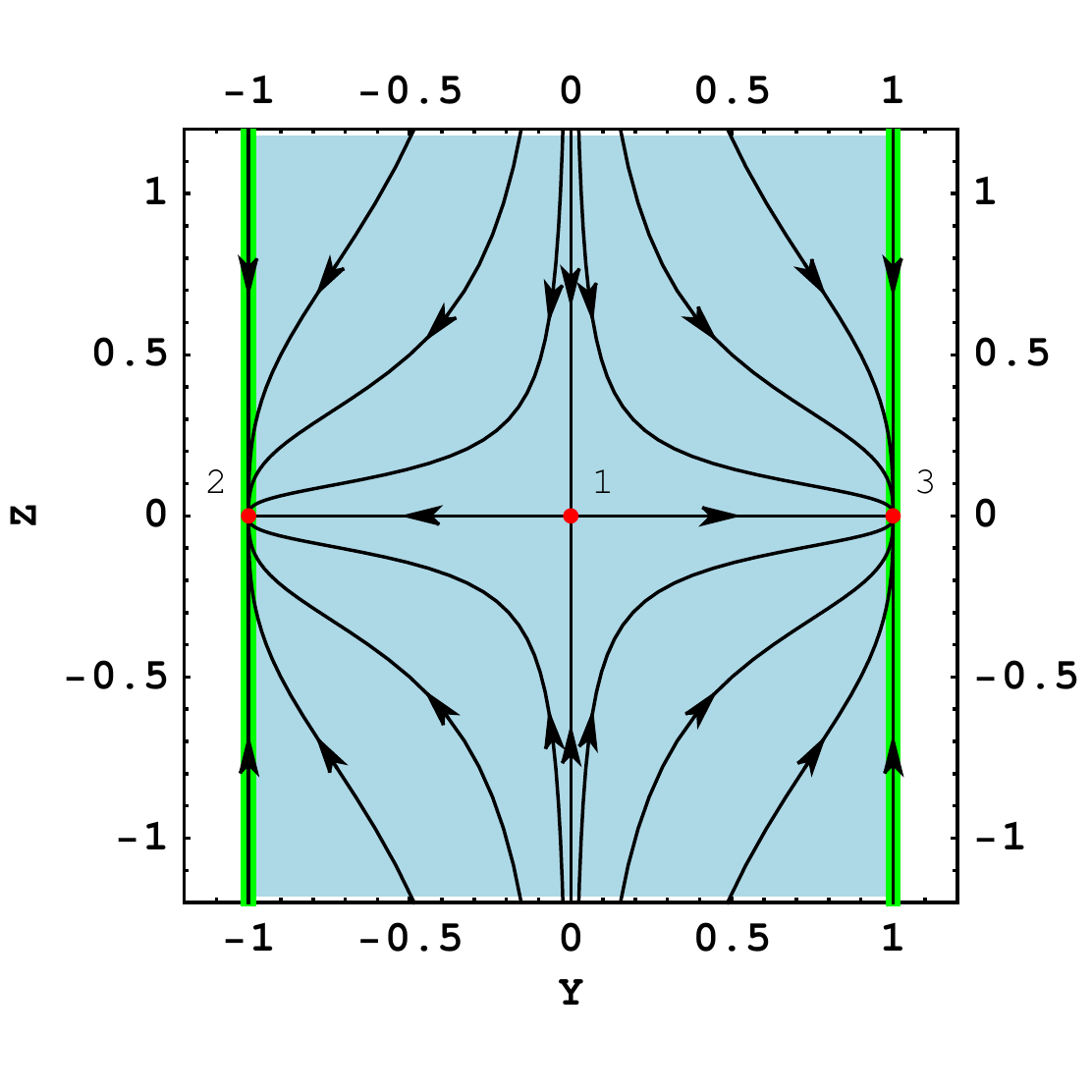}
	\caption{The phase portrait of the invariant submanifold $X=0$ of the dynamical system (\ref{dyn3u})-(\ref{dyn4u}). Critical point (1) represents a Einstein-de Sitter universe. Critical points (3) represents a stable de Sitter. Critical points (2) represents a contracting de Sitter universe. Note that because of time parametrization $dt = Hd\tau$ in the region $X < 0$, the cosmological time $t$ is reversed. In consequence critical point (2) is unstable. The de Sitter universe is located on the invariant submanifold $\{ \frac{\dot{H}}{H^2}=0 \}$ which is represented by green vertical lines. By identification of green lines of the phase portrait one can represent the dynamics on the cylinder. The boundary of the physical region is restricted with the condition $Y^2\le 1$, which is a consequence $\Omega_{\text{m}} \ge 0$. Note that, trajectories achieve the de Sitter states along tangential vertical lines.}
	\label{fig:23}
\end{figure}

The solution of dynamical system (\ref{dynx1})-(\ref{dynx2}) is $x=C_1 a^{-3}$, $y=\text{const}$ and $z = - \frac{1}{3} C_1 a^{-3}+C_2$.
The phase portraits and a list of critical points for dynamical system (\ref{dyn3y})-(\ref{dyn4y}) is presented in Figures \ref{fig:8}, \ref{fig:18} and Table \ref{tab:4}, respectively. For illustration the behaviour of the trajectories near the invariant submanifold (represented by green lines) in the phase portrait (\ref{fig:19}) we construct two-dimensional phase portraits \ref{fig:23}. In the latter trajectories reach the stationary states along tangential vertical lines (green lines). Critical point (1) is representing the matter dominating universe---an Einstein-de Sitter universe.

\begin{table}[htb]
	\caption{The complete list of critical points of the autonomous dynamical system (\ref{dyn3y})-(\ref{dyn2y}) which are shown in Fig.~\ref{fig:8} and \ref{fig:18}. Coordinates, eigenvalues of the critical point as well as its type and cosmological interpretation are given.}.
	\label{tab:4}
	\begin{tabular}{ccccc}
		\hline Critical point & Coordinates & Eigenvalues & Type of critical point & Type of universe\\ \hline
		1 & $x=0$, $y=0$ & 3 -3 &  saddle & Einstein-de Sitter\\
		2 & $x=0$, $y=-1$ & -3, -3 &  stable node & contracting de Sitter\\
		3 & $x=0$, $y=1$& -3, -3 &  stable node & de Sitter\\
		4 & $x=1$, $y=0$ & 3, 3 &  unstable node & Zeldovitch stiff\\
		  &              &      &                & matter dominating\\
		5 & $x=-1$, $y=0$& 3,  3 &  unstable node & Zeldovitch stiff\\
		  &              &       &                & matter dominating\\
		\hline
	\end{tabular}
\end{table}

Finally $\Omega_{\Lambda,\text{emergent}}$ for first case is in the following form
\begin{equation}
\Omega_{\Lambda,\text{emergent}}=\Omega_{\Lambda,\text{emergent,0}}a^{-6}+\Omega_{\Lambda,\text{0}}.
\end{equation}

\begin{figure}[ht]
	\centering
	\includegraphics[width=0.7\linewidth]{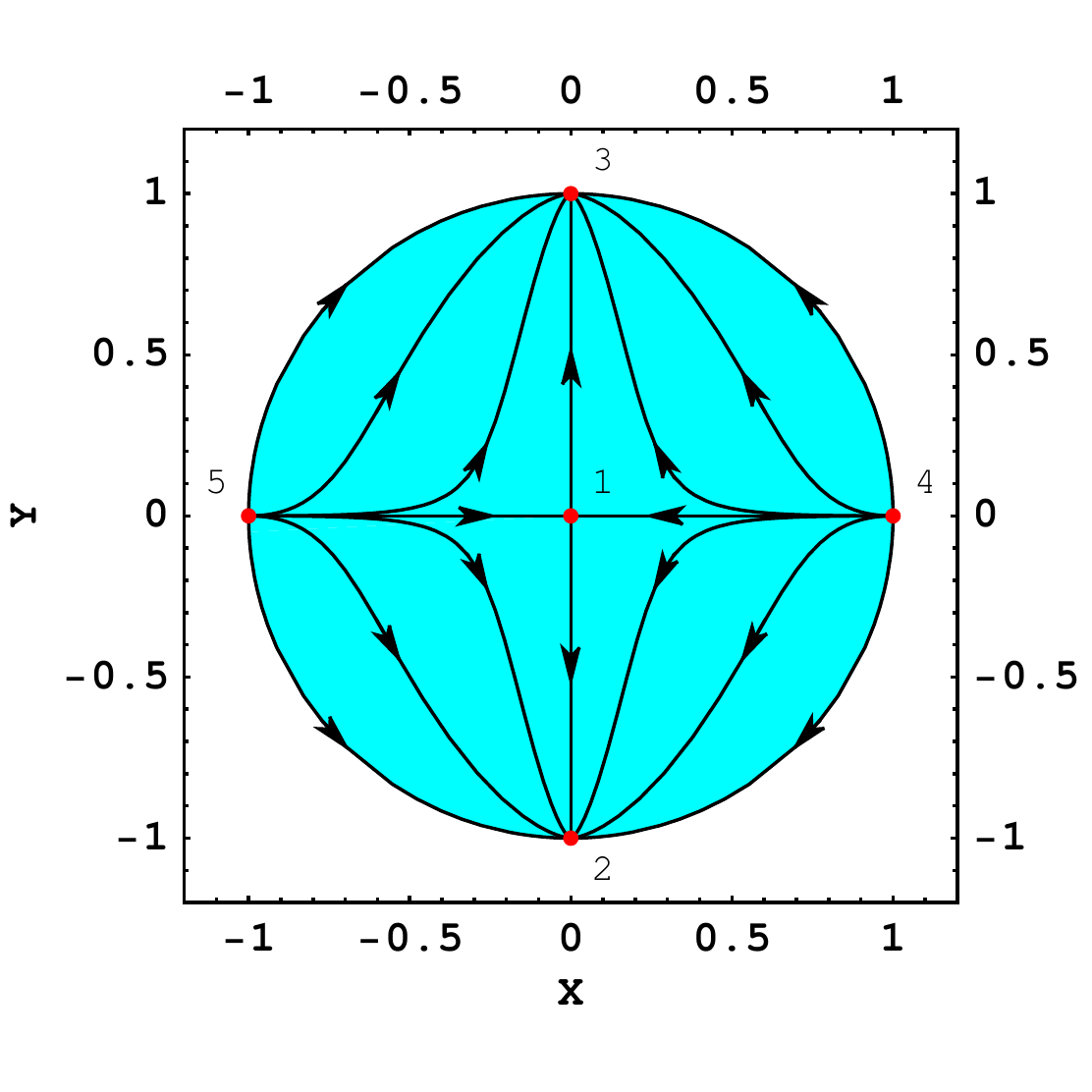}
	\caption{The phase portrait for dynamical system (\ref{dyn3u})-(\ref{dyn5u}). Critical points (1) represents a Einstein-de Sitter universe. Critical point (4) and (5) represent a Zeldovitch stiff matter universe. Critical points (2) represents a contracting  de Sitter. Critical points (3) represents stable de Sitter universe. The de Sitter universe is located on the invariant submanifold $\frac{\dot{H}}{H^2}=0$. The blue region presents physical restricted by condition $X^2+Y^2\le 1$, which is a consequence of $\Omega_{\text{m}} \ge 0$.}
	\label{fig:1}
\end{figure}

Now, let us concentrate on the second case. The system (\ref{dyn3})-(\ref{dyn4}) assumes the following form
\begin{align}
\frac{dx}{d\tau}&=-3x-2z,\label{dyn5}\\
\frac{dy}{d\tau}&=0 \quad \Rightarrow \quad y=\text{const},\\
\frac{dz}{d\tau}&=x\label{dyn6}
\end{align}
with condition
\begin{equation}
0=(x+z)^2-3y^2+3.\label{con2}
\end{equation}
The dynamical system  (\ref{dyn5})-(\ref{dyn6}) is linear and can be simply integrated. The solution of above equations are $x= -2C_1 a^{-2}-C_2 a^{-1}$, $y=\text{const}$ and $z=C_1 a^{-2}+C_2 a^{-1}$.
The phase portrait and critical points for dynamical system (\ref{dyn3})-(\ref{dyn4}) is presented in Figures \ref{fig:1}, \ref{fig:19} and Table \ref{tab:3}.

\begin{table}[htb]
	\caption{The list of critical points for the autonomous dynamical system (\ref{dyn3u})-(\ref{dyn5u}) which are shown in Fig.~\ref{fig:1} and \ref{fig:19}. Coordinates, eigenvalues of the critical point as well as its type and cosmological interpretation are given.}.
	\label{tab:3}
\begin{tabular}{ccccc}
	\hline Critical point & Coordinates & Eigenvalues & Type of critical point & Type of universe\\ \hline
	1 & $X=0$, $Y=0$ & 3/2, -1/2 &  saddle & Einstein-de Sitter\\
	2 & $X=0$, $Y=-1$ & -3, -2 &  stable node & contracting de Sitter\\
	3 & $X=0$, $Y=1$ & -3, -2 &  stable node & de Sitter\\
	4 & $X=1$, $Y=0$ & 1, 2 &  unstable node & Zeldovitch stiff\\
	  &              &      &                & matter dominating\\
	5 & $X=-1$, $Y=0$ & 1,  2 &  unstable node & Zeldovitch stiff\\
	  &               &       &                & matter dominating\\
	\hline
\end{tabular}
\end{table}

On the invariant submanifold (\ref{con2}) dynamical system (\ref{dyn5})-(\ref{dyn6}) reduces to
\begin{align}
\frac{dx}{d\tau}&=-x,\label{dyn5xx}\\
\frac{dz}{d\tau}&=-z.\label{dyn6xx}
\end{align}
The solutions of (\ref{dyn5xx})-(\ref{dyn6xx}) are $x=C_1 a^{-1}$ and $z=C_2 a^{-1}$.

Finally
\begin{equation}
\Omega_{\Lambda,\text{emergent}}=\Omega_{\Lambda,0}+\Omega_{\Lambda,\text{emergent},0}a^{-4},
\end{equation}
i.e., relation $\Lambda(a)\propto a^{-4}$ arises if we consider the behaviour of trajectories in the neighborhood of an unstable de Sitter state $\frac{\dot{H}}{H^2}=0$.
Therefore the emergent term is of the type `radiation'. In the scalar field cosmology there is phase of evolution during which effective coefficient e.o.s. is 1/3 like for radiation. If we have find trajectory in the neighborhood of a saddle point then such a type of behavior appears \cite{Szydlowski:2008in}.

\section{How to constraint emergent running $\Lambda(a)$ cosmologies?}
Dark energy can be divided into two classes: with or without early dark energy \cite{Pettorino:2013ia}. Models without early dark energy behave like the $\Lambda$CDM model in the early time universe. For models with early dark energy, dark energy has an important role in evolution of early universe. These second type models should have a scaling or attractor solution where the fraction of dark energy follows the fraction of the dominant matter or radiation component. In this case, we use fractional early dark energy parameter $\Omega^{\text{e}}_\text{d}$ to measured ratio of dark energy to matter or radiation.

The model with with $\xi=1/6$ (conformal coupling) and $V=\text{const}$ belongs to the class of models with early constant ratio dark energy in which $\Omega_{\text{de}}$=const during the radiation dominated stage. In this case we can use the fractional early dark energy parameter $\Omega^{\text{e}}_{\text{d}}$ \cite{Doran:2006kp,Pettorino:2013ia} which is constant for models with constant dark energy in the early universe. The fractional density of early dark energy is defined by the expression $\Omega_{\text{d}}^{\text{e}}=1-\frac{\Omega_{\text{m}}}{\Omega_{\text{tot}}}$, where $\Omega_{\text{tot}}$ is the sum of dimensionless density of matter and dark energy. In this case, there exist strong observational upper limits on this quantity \cite{Ade:2013zuv}.

For this aim let us note that during the `radiation' epoch we can apply this limit $\Omega_{\text{d}}^{\text{e}}<0.0036$ \cite{Ade:2013zuv} and
\begin{equation}
1-\Omega_{\text{d}}^{\text{e}}=\frac{\Omega_{\text{m},0}a^{-3}+\Omega_{\text{r,0}}a^{-4}}{\Omega_{\text{m,0}}a^{-3}+\Omega_{\text{r,0}}a^{-4}+\Omega_{\Lambda,0}+\Omega_{\text{em,0}}a^{-4}}.
\end{equation}

Let us consider radiation dominating phase $a(t)\propto t^{\frac{1}{2}}$ ($p_{\text{eff}}=\frac{1}{3}\rho_{\text{eff}}$) \cite{Szydlowski:2008in}
\begin{equation}
1-\Omega_{\text{d}}^{\text{e}}=\frac{\Omega_{\text{m},0}t^{-\frac{3}{2}}+\Omega_{\text{r},0}t^{-2}}{\Omega_{\text{m},0}t^{-\frac{3}{2}}+\Omega_{\text{r},0}t^{-2}+\Omega_{\Lambda,0}+\Omega_{\text{em},0}t^{-2}} \backsimeq^{\text{at early universe}} \frac{\Omega_{\text{r},0}}{\Omega_{\text{r},0}+\Omega_{\text{em},0}}.
\end{equation}

$\Omega_{\text{d}}^{\text{e}}$ at the early universe is constant and
\begin{equation}
\Omega_{\text{d}}^{\text{e}}=\frac{\Omega_{\text{r},0}}{\Omega_{\text{r},0}+\Omega_{\text{em},0}}<0.0036.
\end{equation}
From the above formula we get that $\frac{\Omega_{\text{em,0}}}{\Omega_{\text{r,0}}}<0.003613$. In consequence we have a stricter limit on a strength of the running $\Lambda$ parameter in the present epoch $\Omega_{\text{em},0}<3.19\times 10^{-7}$.

\section{Cosmology with non-canonical scalar field}
The dark energy can be also parametrized in a covariant way by a no-canonical scalar field $\phi$ \cite{Sahni:2015hbf}. The main difference between canonical and non-canonical description of scalar field is in the generalized form of the pressure $p_{\phi}$ of the scalar field. For the canonical scalar field, the pressure $p_{\phi}$ is expressed by the formula $p_{\phi}=\frac{\dot\phi^{2}}{2}-V(\phi)$, where ${ }^{\dot{ }}\equiv\frac{d}{dt}$ and $V(\phi)$ is the potential of the scalar field. In the non-canonical case, the pressure is described by the expression $p_{\phi}=\left(\frac{\dot\phi^{2}}{2}\right)^{\alpha}-V(\phi)$, where $\alpha$ is an additional parameter. If $\alpha$ is equal 1 then the pressure of the non-canonical scalar field represents the canonical case.

\begin{figure}
	\centering
	\includegraphics[width=0.7\linewidth]{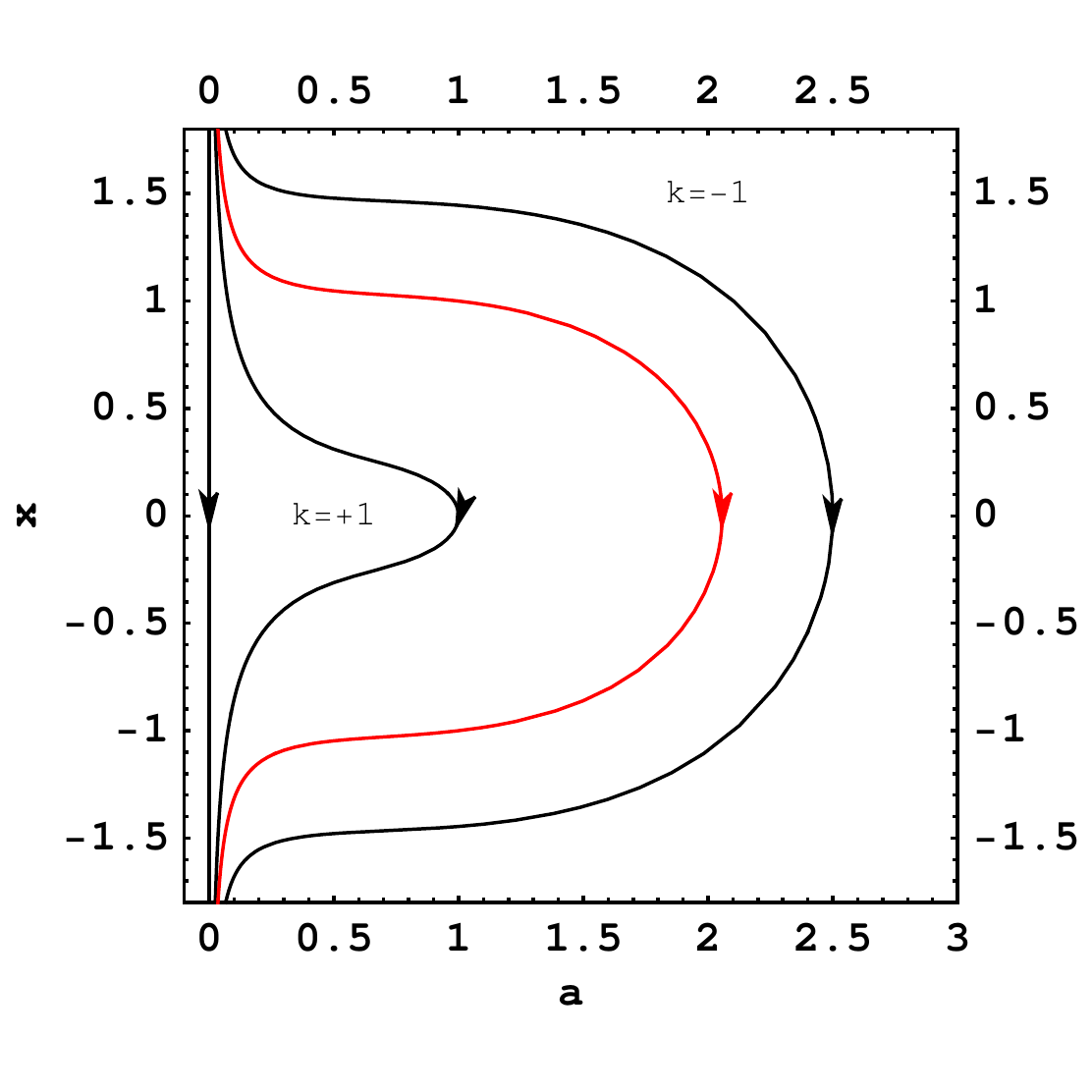}
	\caption{A phase portrait for dynamical system (\ref{noncana})-(\ref{noncanx}) for example with $\alpha=1/8$. Red lines represent the flat universe and these trajectories separates the region in which lies closed and open models. Note that all models independence on curvature are oscilating.}
	\label{fig:12}
\end{figure}

\begin{figure}
	\centering
	\includegraphics[width=0.7\linewidth]{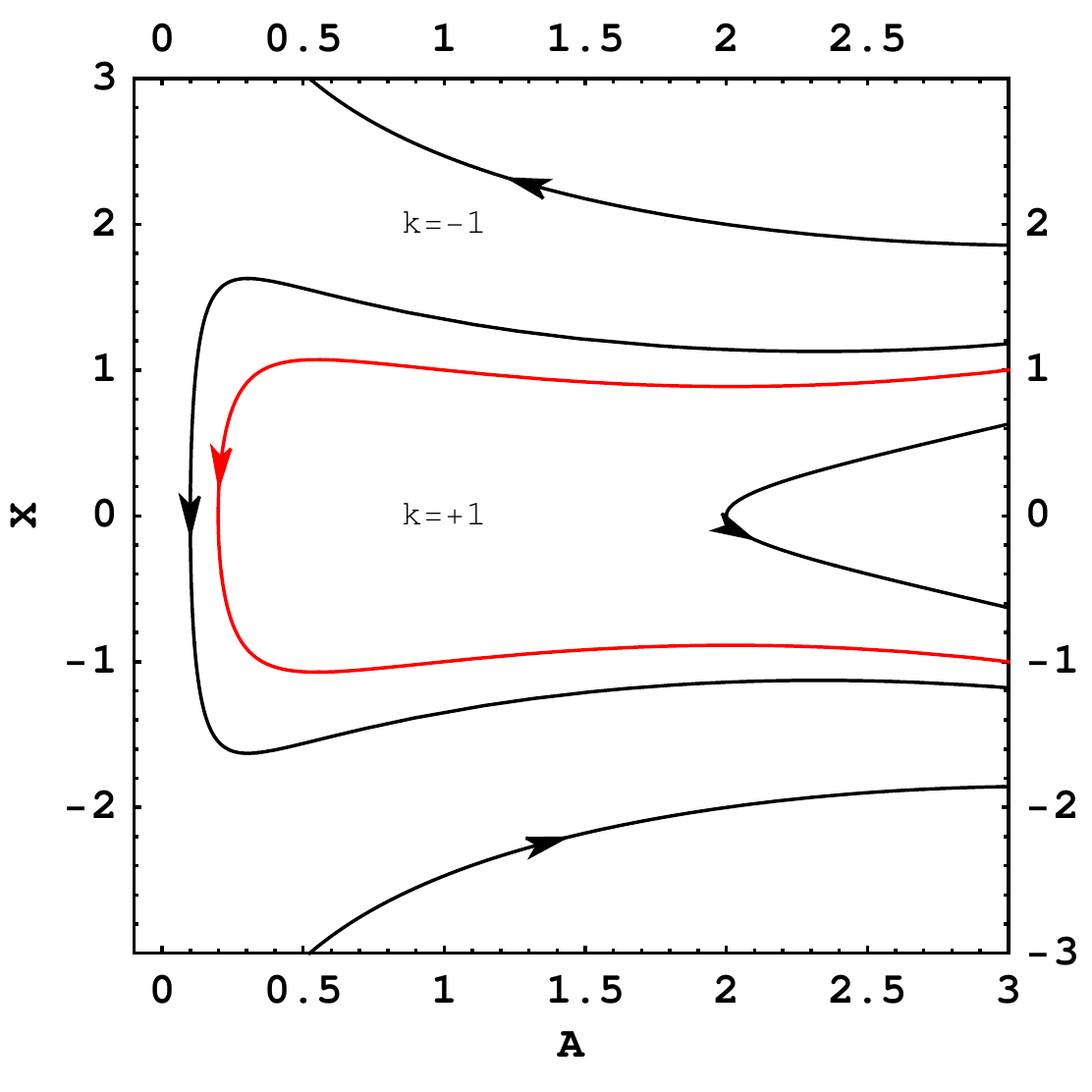}
	\caption{A phase portrait for dynamical system (\ref{noncanA})-(\ref{noncanX}) for example with $\alpha=1/8$. Red lines represent the flat universe and these trajectories separates the region in which lies closed and open models.}
	\label{fig:13}
\end{figure}

\begin{figure}
	\centering
	\includegraphics[width=0.7\linewidth]{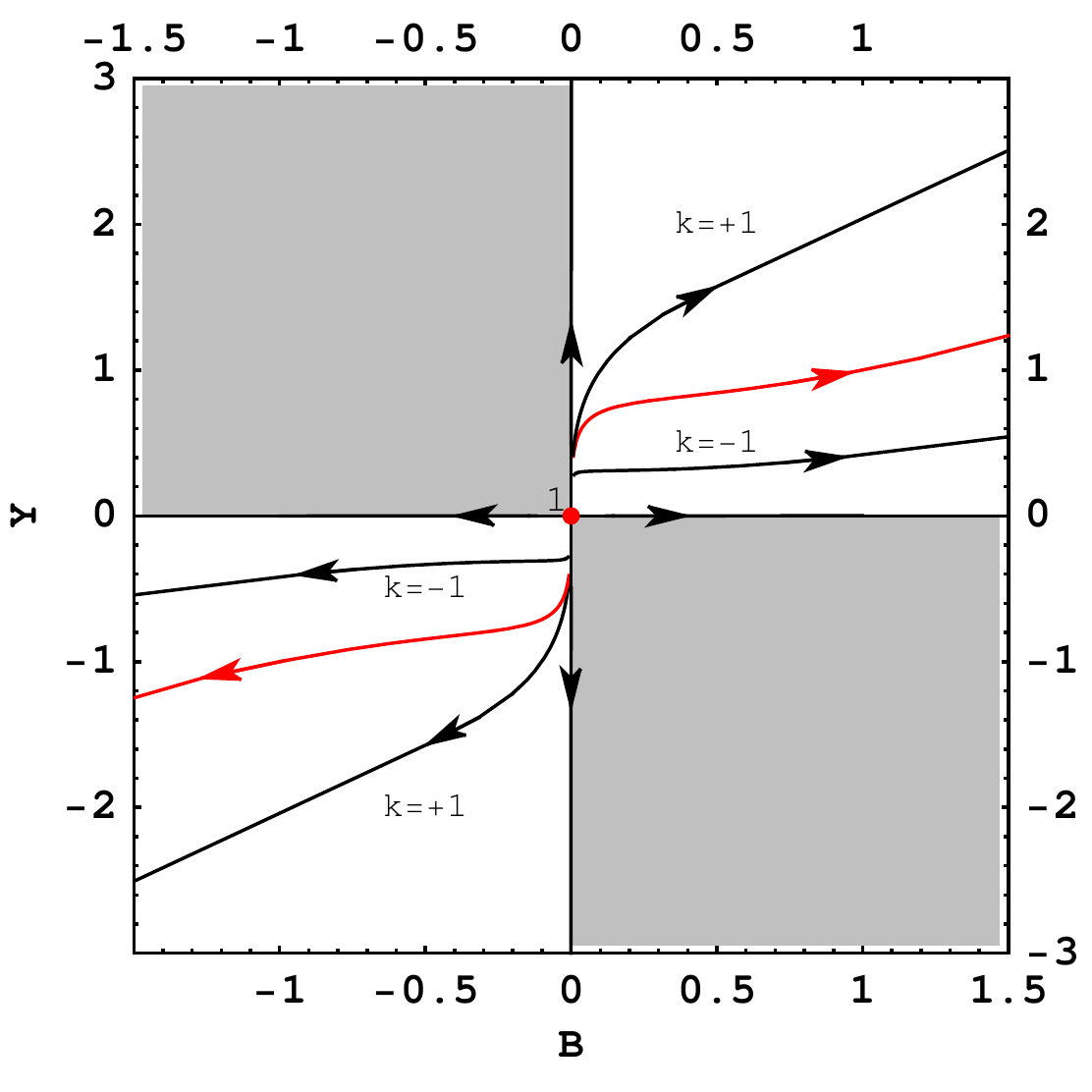}
	\caption{A phase portrait for dynamical system (\ref{noncanB})-(\ref{noncanY}) for example with $\alpha=1/8$. Critical point (1) at the origin $B=0$, $Y=0$ presents stable node and Einstein-de Sitter universe. Grey region represents non-physical domain excluded by the condition $\tilde{X}\tilde{Y}>0$. Red lines represent the flat universe and these trajectories separates the region in which lies closed and open models.}
	\label{fig:14}
\end{figure}

\begin{figure}
	\centering
	\includegraphics[width=0.7\linewidth]{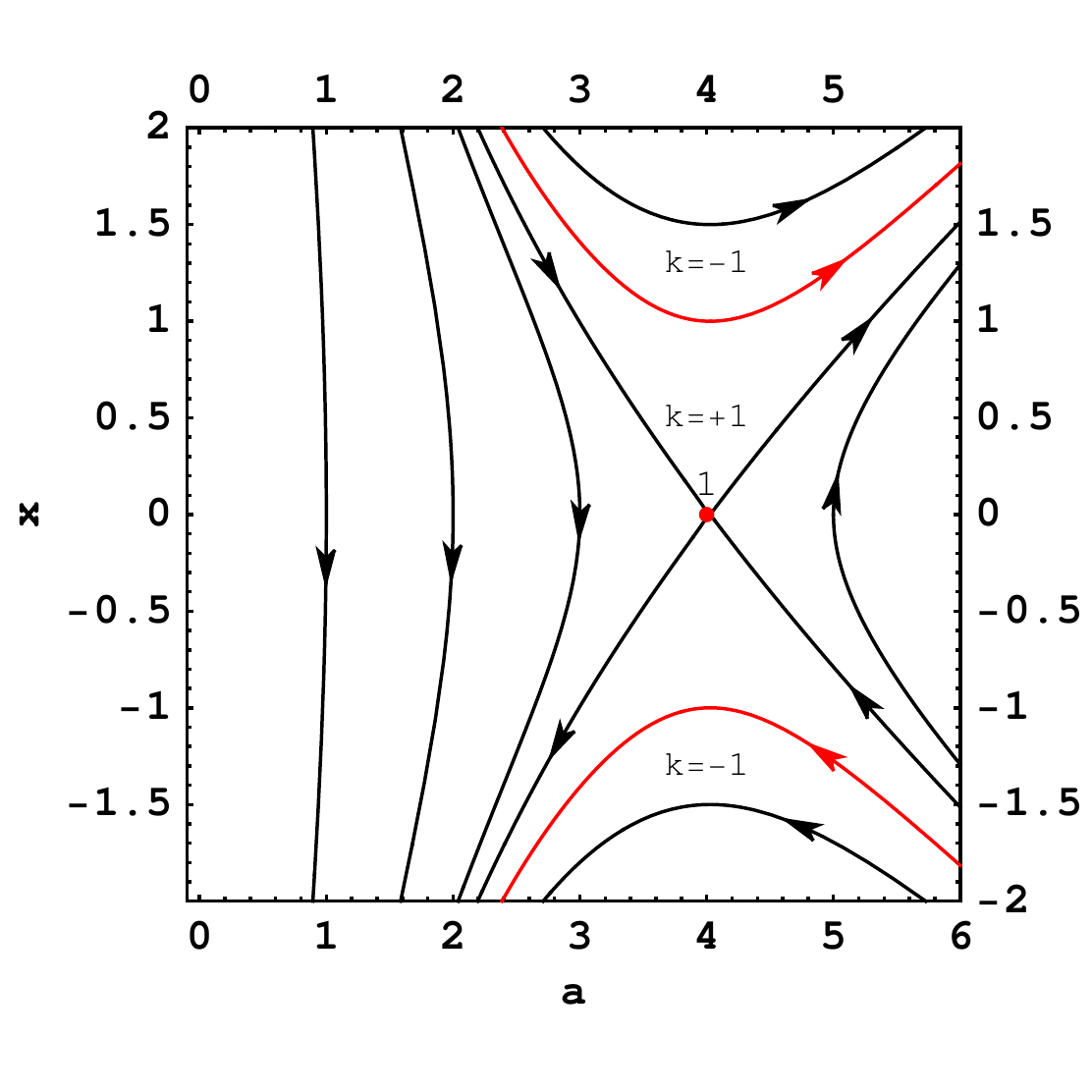}
	\caption{A phase portrait for dynamical system (\ref{noncana})-(\ref{noncanx}) for example with $\alpha=50$.  Critical point (1) presents saddle and represents static Einstein universe. Red lines represent the flat universe and these trajectories separates the region in which lies closed and open models. Note that all models independence on curvature are oscilating.}
	\label{fig:20}
\end{figure}

\begin{figure}
	\centering
	\includegraphics[width=0.7\linewidth]{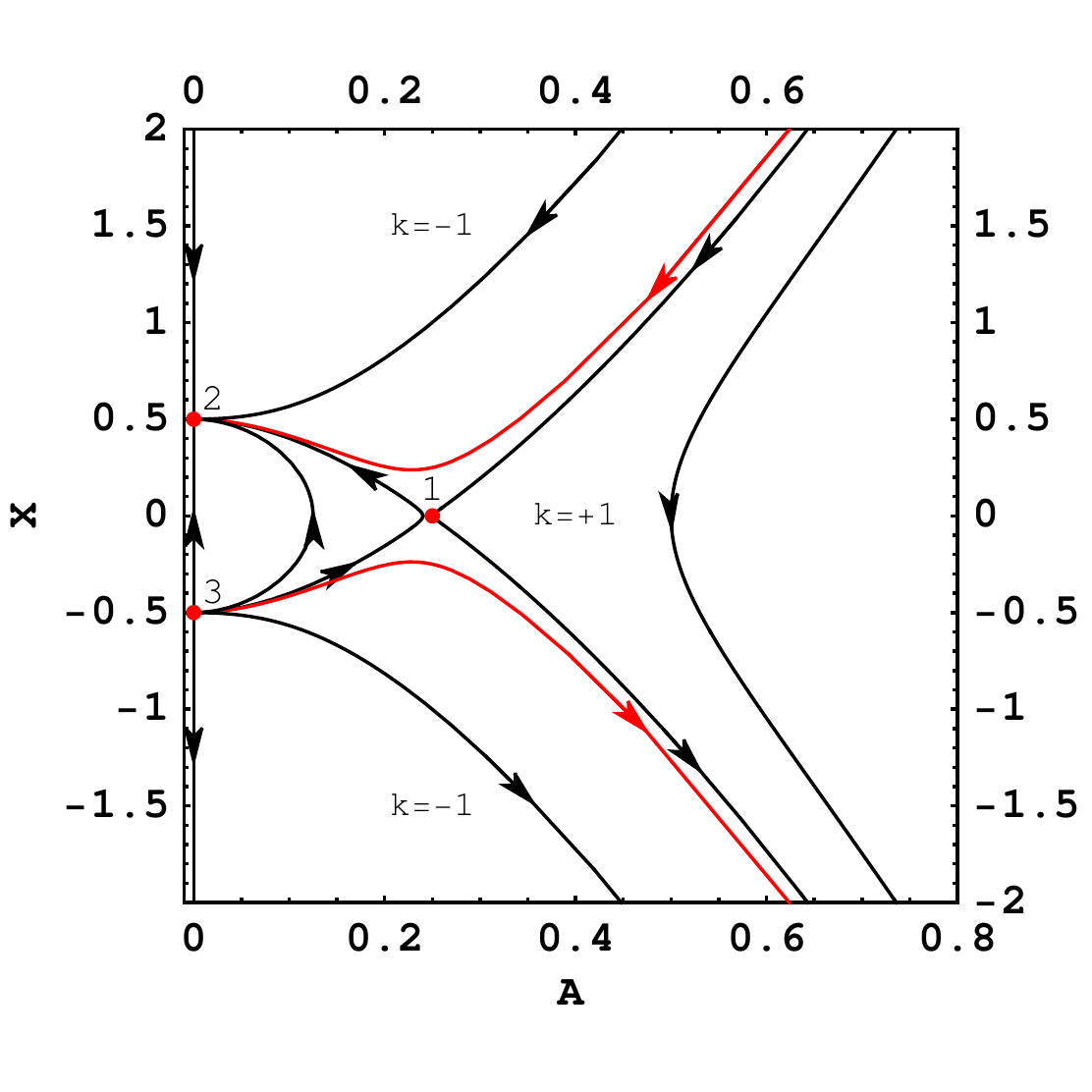}
	\caption{A phase portrait for dynamical system (\ref{noncanA})-(\ref{noncanX}) for example with $\alpha=50$. Critical point (1) presents saddle and represents static Einstein universe. Critical points (2) represents a stable de Sitter. Critical points (3) represents a contracting de Sitter universe. Red lines represent the flat universe and these trajectories separates the region in which lies closed and open models.}
	\label{fig:21}
\end{figure}

\begin{figure}
	\centering
	\includegraphics[width=0.7\linewidth]{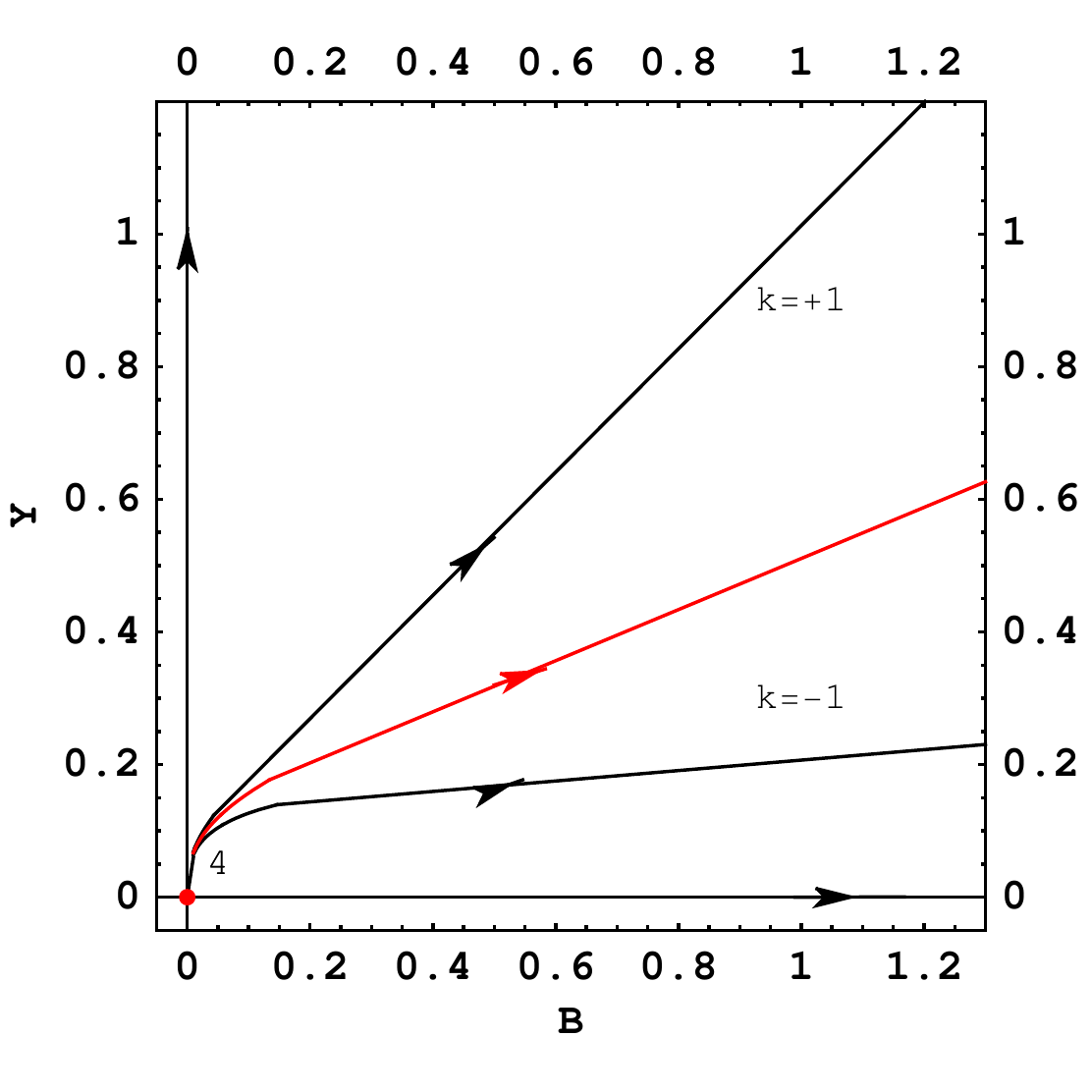}
	\caption{A phase portrait for dynamical system (\ref{noncanB})-(\ref{noncanY}) for example with $\alpha=50$. Critical point (4) at the origin $B=0$, $Y=0$ presents an unstable node and an Einstein-de Sitter universe. Red lines represent the flat universe and these trajectories separates the region in which lies closed and open models.}
	\label{fig:22}
\end{figure}

The theory of the non-canonical scalar field is of course covariant formulation because this theory can be obtain from the action, which is described by the following formula
\begin{equation}
S=\int\sqrt{-g}\left(R+\left(\frac{\dot\phi^{2}}{2}\right)^{\alpha}-V(\phi)+\mathcal{L}_{\text{m}}\right)d^4 x,
\end{equation}
where $\mathcal{L}_{\text{m}}$ is the Lagrangian for the matter.
After variation of the Lagrangian $\mathcal{L}$ with respect the metric we get Friedmann equations in the following form
\begin{equation}
3H^2=\rho_{\text{m}}+(2\alpha-1)\left(\frac{\dot\phi^{2}}{2}\right)^{\alpha}+V(\phi)-\frac{3 k}{a^{2}},\label{friedmann1}
\end{equation}
\begin{equation}
-3\frac{\ddot a}{a}=\frac{\rho_{\text{m}}}{2}+(\alpha-1)\left(\frac{\dot\phi^{2}}{2}\right)^{\alpha}-V(\phi).\label{friedmann2}
\end{equation}

We obtain an additional equation of motion for a scalar field after variation of the Lagrangian $\mathcal{L}$ with respect the scalar field
\begin{equation}
\ddot{\phi}+\frac{3H\dot\phi}{2\alpha-1}+\left(\frac{V'(\phi)}{\alpha(2\alpha-1)}\right)\left(\frac{2}{\dot\phi^2}\right)^{\alpha-1}=0,\label{klein}
\end{equation}
where $'\equiv\frac{d}{d\phi}$.

For $\alpha=1$, equations (\ref{friedmann1}), (\ref{friedmann2}) and (\ref{klein}) reduce to the case of the canonical scalar field. For $\alpha=0$ we have the case with the constant scalar field. It is an interesting case with $\alpha=2$ and the constant potential $V$. There, the scalar field imitates radiation because the $\phi^{2\alpha}\propto a^{-4}$ in the Friedmann equation.

For the constant potential $V=\Lambda$ equation (\ref{klein}) reduces to
\begin{equation}
\ddot{\phi}+\frac{3H\dot\phi}{2\alpha-1}=0.\label{klein2}
\end{equation}
The above equation has the following solution
\begin{equation}
\dot{\phi}=\phi_0 a^{\frac{-3}{2\alpha-1}}.\label{field}
\end{equation}

We can obtain from (\ref{friedmann1}), (\ref{friedmann2}) and (\ref{klein}) the dynamical system for the non-canonical scalar field with the constant potential in variables $a$ and $x=\dot a$
\begin{equation}
a'=x a^2,\label{noncana}
\end{equation}
\begin{equation}
x'=-\frac{\rho_{\text{\text{m},0}}}{6}-\frac{\alpha+1}{3}a^{\frac{3}{1-2\alpha}}+\frac{\Lambda}{3}a^3,\label{noncanx}
\end{equation}
where $'\equiv a^2 \frac{d}{dt}$.
The phase portrait for above dynamical system are presented in Figures \ref{fig:12} and \ref{fig:20}.

System (\ref{noncana})-(\ref{noncanx}) possesses critical points only two types:
\begin{enumerate}
	\item static critical points $x_0 =0$,
	\item non-static critical points: $a_0=0$ (Big Bang singularity).
\end{enumerate}

If we assume the matter in the form of dust ($p=0$) then non-static critical points cannot exist at finite domain of phase space. The Big Bang singularity is correspoding with critical points at infinity.

Static critical points lie on the $a$-axis. Linearization of the system in the vinicity of static critical points is establish by linearization matrix $\mathbf{A}$
\begin{equation}
\mathbf{A}=\left(\begin{array}{cc}
2x a & a^{2}\\
3\frac{1+\alpha}{-1+2\alpha}a^{2\left(\frac{\alpha+1}{1-2\alpha}\right)}+3\Lambda a^2 & 0
	\end{array}\right)\Big|_{(a_0,0)},
	\end{equation}
	where $a_0$ is the solution of the algebraic equation
	\begin{equation}
	2\Lambda a^3=\rho_{\text{m,0}}+2(\alpha+1)a^{\frac{3}{1-2\alpha}}.
	\end{equation}
Hence $\tr A = 0$ and
\begin{equation}
\det (\mathbf{A}-\mu \mathds{1} )=3\frac{(1+\alpha)}{{1-2\alpha}}a^{2\left(\frac{2-\alpha}{1-2\alpha}\right)}-3\Lambda a^4-2x a\mu+\mu^2,
\end{equation}

where $\mu$ is eigenvalue. Therefore the characteristic equation, for the critical point $(a_0,0)$ assumes the simple form
\begin{equation}
\mu^2=3\Lambda a_0^4-3\frac{(1+\alpha)}{{1-2\alpha}}a_0^{2\left(\frac{2-\alpha}{1-2\alpha}\right)}
\end{equation}

Note that, if $\alpha>\frac{1}{2}$ then eigenvalues for critical point ($a_0,0$) are real and correspond with a saddle type of critical point. Therefore for $\alpha>\frac{1}{2}$ the qualitative structure of the phase space is topologically equivalent (by homeomorphism) of the $\Lambda$CDM model. Hence, the phase space portrait is structural stable, i.e. it is not disturbed under small changes of its right-hand sides.

For the analysis the behaviour of trajectories in the infinity we use the following sets of projective coordinates:
\begin{enumerate}
	\item $A=\frac{1}{a}$, $X=\frac{x}{a}$,
	\item $B=\frac{a}{x}$, $Y=\frac{1}{x}$.
\end{enumerate}
Two maps cover the behaviour of trajectories at the circle at infinity.

The dynamical system for variables $A$ and $X$ is expressed by
\begin{align}
A'&=-XA^2,\label{noncanA}\\
X'&=A^{4}\left(-\frac{\rho_{\text{\text{m},0}}}{6}-\frac{\alpha+1}{3}A^{\frac{3}{2\alpha-1}}\right)+A\left(\frac{\Lambda}{3}-X^2\right),\label{noncanX}
\end{align}
where $'\equiv A \frac{d}{dt}$.
The dynamical system for variables $B$ and $Y$ is expressed by
\begin{align}
\dot B&=BY\left(B+\left(\frac{\rho}{6}Y^3+\frac{\alpha+1}{3}B^{\frac{3}{1-2\alpha}}Y^{\frac{6\alpha}{2\alpha-1}}-\frac{\Lambda}{3}B^3\right)\right).\label{noncanB}\\
\dot Y&=Y^2\left(\frac{\rho}{6}Y^3+\frac{\alpha+1}{3}B^{\frac{3}{1-2\alpha}}Y^{\frac{6\alpha}{2\alpha-1}}-\frac{\Lambda}{3}B^3\right),\label{noncanY}
\end{align}
where $'\equiv B^2 Y \frac{d}{dt}$.

From the analysis of above dynamical systems we found one critical point ($B=0$, $Y=0$) which represents the Einstein-de Sitter universe. Phase portraits for above dynamical systems are presented Figures \ref{fig:13}, \ref{fig:14}, \ref{fig:22} and \ref{fig:21}.

\section{Cosmology with diffusion}
The parametrization of dark energy can be also described in terms of scalar field $\phi$ ~\cite{Boehmer:2015sha,Calogero:2013zba}. As an example of such a covariant parametrization of $\Lambda$ let us consider the case of cosmological models with diffusion. In this case the Einstein equations and equations of current density $J^{\mu}$ are the following
\begin{equation}
R_{\mu\nu}-\frac{1}{2}g_{\mu\nu}R+\phi g_{\mu\nu}=T_{\mu\nu},\label{dif1}
\end{equation}
\begin{equation}
\nabla_{\mu}T^{\mu\nu}=\sigma J^{\nu},\label{dif2}
\end{equation}
\begin{equation}
\nabla_{\mu}J^{\mu}=0,\label{dif5}
\end{equation}
where $\sigma$ is a positive parameter.

From the Bianchi identity $\nabla^{\mu}\left(R_{\mu\nu}-\frac{1}{2}g_{\mu\nu}R\right)=0$, equation (\ref{dif1}) and (\ref{dif2}) we get the following expression for $\Lambda(a(t))$
\begin{equation}
\nabla_\mu \phi=\sigma J_{\mu}.\label{dif4}
\end{equation}
We assume also that the matter is a perfect fluid. Then the energy-momentum tensor is expressed in the following form
\begin{equation}
T_{\mu\nu}=\rho u_{\mu}u_{\nu}+p\left(g_{\mu\nu}+u_{\mu}u_{\nu}\right),
\end{equation}
where $u_{\mu}$ is the 4-velocity and current density is expressed by
\begin{equation}
J^{\mu}=nu^{\mu}.
\end{equation}

\begin{figure}
	\centering
	\includegraphics[width=0.7\linewidth]{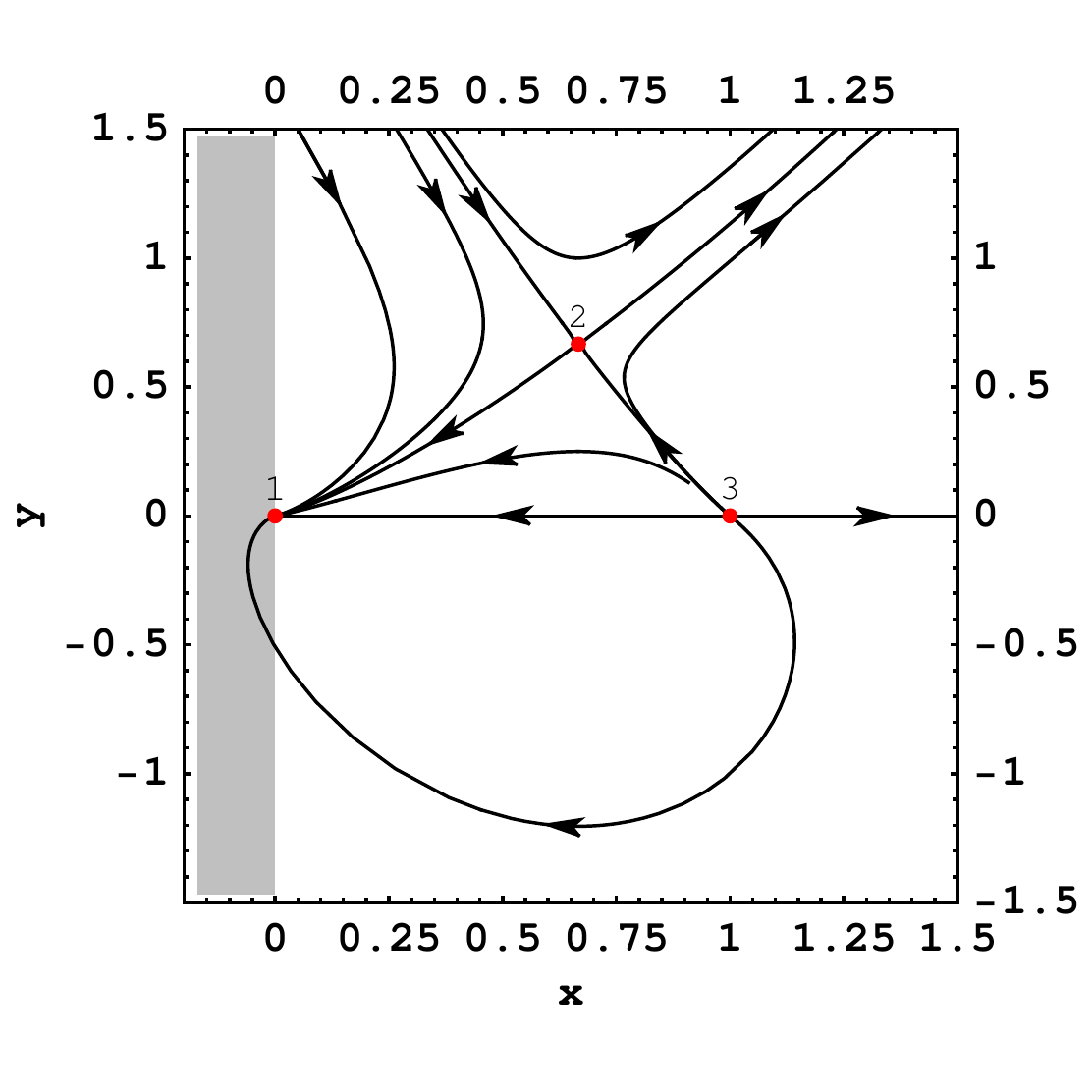}
	\caption{A phase portrait for dynamical system (\ref{dyfdyn1})-(\ref{dyfdyn1}). Critical point (1) ($x=0$, $y=0$) is stable and is in the node type and represents de Sitter universe. Critical point (2) ($x=2/3$, $y=2/3$) represents the saddle type and the static Einstein universe. Critical point (3) ($x=1$, $y=0$) is unstable and is in the node type and represents the Einstein-de Sitter universe. Note the existence of pathological trajectories crossing over the boundary $x=\rho_{\text{m}}=0$}
	\label{fig:15}
\end{figure}

\begin{figure}
	\centering
	\includegraphics[width=0.7\linewidth]{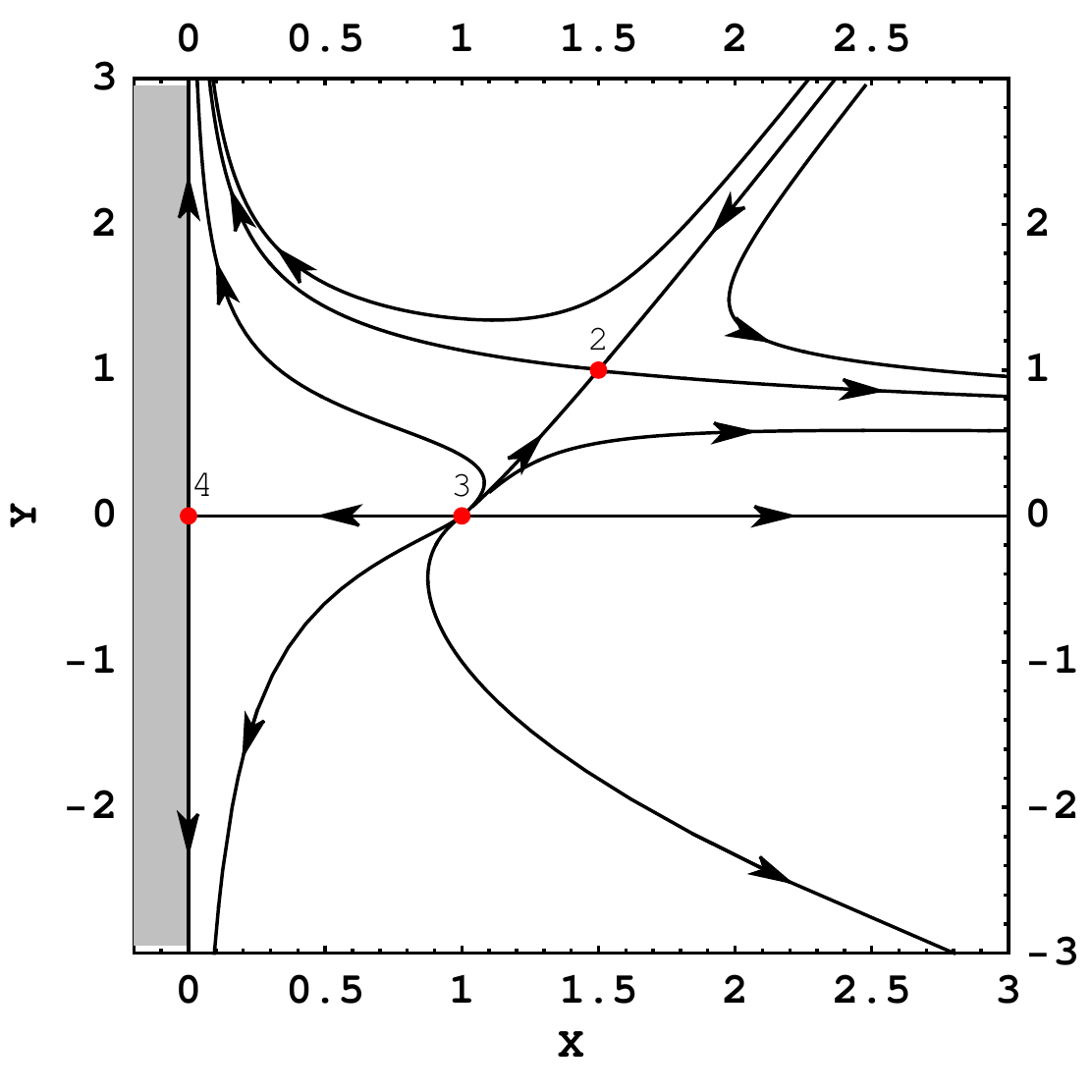}
	\caption{A phase portrait for dynamical system (\ref{dyfdyn31x})-(\ref{dyfdyn33x}). Critical point (4) ($X=0$, $Y=0$) is saddle type and represents the Einstein-de Sitter universe. Critical point (2) ($X=3/2$, $Y=1$) is a saddle type and represents the static Einstein universe.}
	\label{fig:16}
\end{figure}

\begin{figure}
	\centering
	\includegraphics[width=0.7\linewidth]{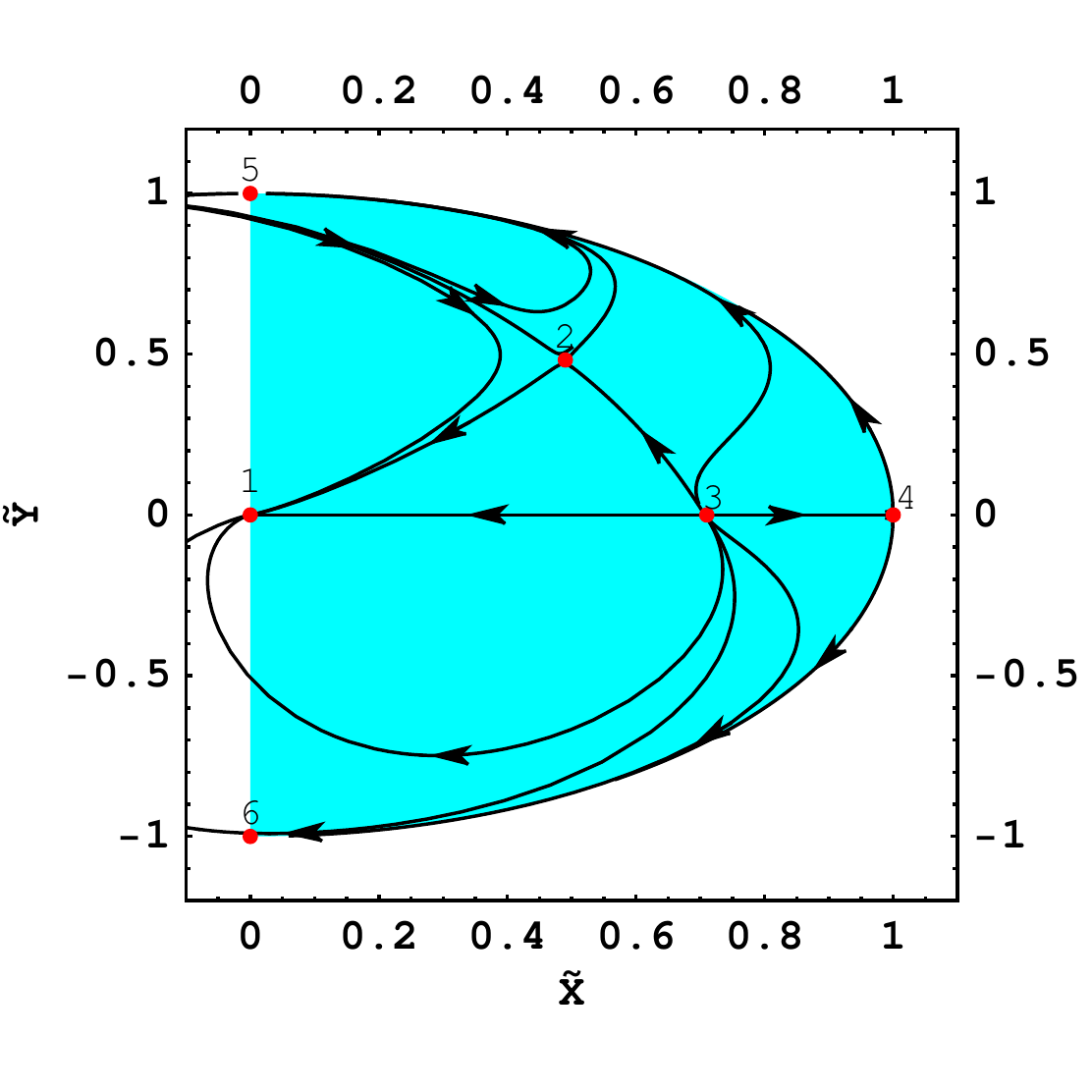}
	\caption{A phase portrait for dynamical system (\ref{dyfdyn31y})-(\ref{dyfdyn33y}). The blue region represents physical domain. Critical points (5) an (6) ($\tilde{X}=0$, $\tilde{Y}1$ and $\tilde{X}=0$, $\tilde{Y}=-1$) represent the Minkowski universe. The blue region represents a physical domain restricted by $B^2 + Y^2 \le 0$, $B\ge 0$.}
	\label{fig:17}
\end{figure}

Under these considerations equations (\ref{dif2}), (\ref{dif4}) and (\ref{dif5}) are described by the following expressions
\begin{equation}
\nabla_{\mu}(\rho u^{\mu})+p\nabla_{\mu}u^{\mu}=\sigma n,\label{dif3}
\end{equation}
\begin{equation}
\nabla_{\mu}(nu^{\mu})=0,\label{dif6}
\end{equation}
and
\begin{equation}
\nabla_{\mu}\phi=\sigma nu_{\mu}.
\end{equation}

We consider the case cosmological equations for simplicity with the zero curvature. Equation (\ref{dif6}) gets
\begin{equation}
n=n_0 a^{-3}.
\end{equation}
In this case we have the following cosmological equations
\begin{equation}
3H^2=\rho_{\text{m}}+\Lambda(a(t)),
\end{equation}
\begin{equation}
\dot{\rho}_{\text{m}}=-3H\rho_{\text{m}}+\sigma n_0 a^{-3},
\end{equation}
\begin{equation}
\frac{d\phi}{dt}=-\sigma n_0 a^{-3}.
\end{equation}

If we choice dimensionless state variables $x=\frac{\rho_{\text{m}}}{3H^2}$ and $y=\frac{\sigma n_0 a^{-3}}{3H^2}$ and a parametrization of time as $'\equiv\frac{d}{d\ln a}$ then we get the following dynamical system
\begin{equation}
x'=3x(x-1)+y,\label{dyfdyn1}
\end{equation}
\begin{equation}
y'=3y(\frac{3}{2}x-1).\label{dyfdyn2}
\end{equation}
A phase portrait for (\ref{dyfdyn1})-(\ref{dyfdyn2}) is demonstrated in Fig.~\ref{fig:15}.

Dynamical system (\ref{dyfdyn1})-(\ref{dyfdyn2}) can be rewritten in projective variables for the analysis of critical points in the infinity. In this case we use the following projective coordinates: $X=\frac{1}{x}$, $Y=\frac{y}{x}$. For new variables $X$ and $Y$, we obtain
\begin{align}
X'&=X\left(X(3-Y)-3\right),\label{dyfdyn31x}\\
Y'&=Y\left(\frac{3}{2}-XY\right)\label{dyfdyn33x}
\end{align}
where  $'\equiv X\frac{d}{d\ln a}$.

We can use also the Poincar{\'e} sphere to search critical points in the infinity. We introduce the following new variables: $\tilde{X}=\frac{x}{\sqrt{1+x^2+y^2}}$, $\tilde{Y}=\frac{y}{\sqrt{1+x^2+y^2}}$.
In variables $\tilde{X}$, $\tilde{Y}$, we obtain dynamical system in the form
\begin{align}
\tilde{X}'&=(1-\tilde{X}^2)(3\tilde{X}^2+(\tilde{Y}-3\tilde{X})\sqrt{1-\tilde{X}^2-\tilde{Y}^2})-3\tilde{X}\tilde{Y}^2\left(\frac{3}{2}\tilde{X}-\sqrt{1-\tilde{X}^2-\tilde{Y}^2}\right),\label{dyfdyn31y}\\
\tilde{Y}'&=-\tilde{X}\tilde{Y}(3\tilde{X}^2+(\tilde{Y}-3\tilde{X})\sqrt{1-\tilde{X}^2-\tilde{Y}^2})-3(1-\tilde{Y}^2)\tilde{Y}\left(\frac{3}{2}\tilde{X}-\sqrt{1-\tilde{X}^2-\tilde{Y}^2}\right),\label{dyfdyn33y}
\end{align}
where $'\equiv \sqrt{1-\tilde{X}^2-\tilde{Y}^2}\frac{d}{d\ln a}$.
The phase portraits for (\ref{dyfdyn31x})-(\ref{dyfdyn33x}) and (\ref{dyfdyn31y})-(\ref{dyfdyn33y}) are demonstrated in Figures \ref{fig:16} and \ref{fig:17}.

\section{Conclusion}
In this paper we have studied the dynamics of cosmological models with the running cosmological constant term using dynamical system methods. We considered different parametrization of the $\Lambda$ term, which are used in the cosmological applications. The most popular approach is to parametrize the $\Lambda$ term through the scale factor $a$ or Hubble parameter $H$. Cosmological models in which the energy-momentum tensor of matter (we assume dust matter) is not conserved because the interaction between both dark matter and dark energy sectors.

There is a class of parametrization of the $\Lambda$ term through the Ricci scalar (or trace of energy momentum tensor), energy density of the scalar field or their kinetic part, scalar field $\phi$ minimally or non-minimally coupled to gravity. These choices are consistent with the covariance of general relativity.

We have distinguished also a new class of the emergent $\Lambda$ parametrization obtained directly from the exact dynamics, which does not violate the covariance of general relativity. We discovered that the first class of parametrization ($\Lambda(a)$) can be obtain as emergent formulas from th eexact dynamics.

In the consequence energy density behaves deviation from the standard dilution. Due to decaying vacuum energy standard relation $\rho_{\text{m}}\propto a^{-3}$ is modified. From the cosmological point of view these class of models is special case of cosmology with interacting term $Q=-\frac{d\Lambda}{dt}$.

The main motivation of study such models comes from the solution of the cosmological constant problem, i.e., explain why the cosmological upper bound ($\rho_\Lambda\leqslant 10^{-47}$ GeV) dramatically differs from theoretical expectations ($\rho_{\Lambda} \sim 10^{71}$GeV) by more than 100 orders of magnitude \cite{Weinberg:2008co}. In this context the running $\Lambda$ cosmology is some phenomenological approach toward the finding the relation $\Lambda(t)$ lowering the value of cosmological constant during the cosmic evolution.

In the study of the $\Lambda(t)$CDM cosmology different parametrizations of $\Lambda$ term are postulated. Some of them like $\Lambda(\phi)$, $\Lambda(R)$ or $\Lambda(\tr T^\mu_\nu)$, $\Lambda(T)$, where $T=\frac{1}{2}\dot{\phi}^2$ are consistent with principle of covariance of general relativity. Another one like $\Lambda=\Lambda(H)$ are motivated by quantum field theory.

We demonstrated that $\Lambda=\Lambda(a)$ parametrization can be obtain from the exact dynamics of the cosmological models with scalar field and the potential by taking approximation of trajectories in neighborhood of invariant submanifold $\frac{\dot{H}}{H^2}$ of the original system. The trajectories coming toward a stanle de Sitter are mimicking effects of the running $\Lambda(a)$ term. The arbitrary parametrizations of $\Lambda(a)$, in general, violate the covariance of general relativity. However, some of them which emerge from the covariant theory are an effective description of the behaviour of trajectories in the neighborhood of a stable de Sitter state.

In the paper we study in details the cosmological model dynamics in the phase space which is organized by critical points representing stationary states, invariant manifold etc. We study dynamics at finite domain of the phase space as well as at infinity using projective coordinates.

The phase space structure contain all information about dependence of solutions on initial conditions, its stability, genericity etc.

Due to dynamical system analysis we can reveals the physical status of the Alcaniz-Lima ansatz in the $\Lambda(H)$ approach. This solution from the point of view dynamical system theory is universal asymptotics as trajectories are going toward global attractor representing a de Sitter state. In this regime both $\rho_{\Lambda}-\Lambda_{\text{bare}}$ and $\rho_{\text{m}}$ are proportional, i.e., a scaling solution is obvious.

The detailed studies of the dynamics on the phase portraits show us how `large' is the class of running cosmological models for which the concordance $\Lambda$CDM model is a global attractor.

We also demonstrated on the example of cosmological models with non-minimal coupling constant and constant potential that running parts of the $\Lambda$ term can be constrained by Planck data. Applying the idea of constant early dark energy fraction idea and Ade et al. bound we have find a stringent constraint on the value of the running $\Lambda$ term.

In the paper we considered some parametrization of $\Lambda$ term, which violates the covariance of the Lagrangian like $\Lambda(H)$, $\Lambda(a)$ parametrization but is used as a some kind effective description. We observe in the phase space of cosmological models with such a parametrization some pathologies which manifested by trajectories crossing the boundary of line of zero energy density invariant submanifold. It is a consequence of the fact that $\rho_{\text{m}}=0$ is not a trajectory of the dynamical system. On the other hand the $\Lambda(a)$ parametrization can emerge from the basic covariant theory as some approximation of true dynamics.

We illustrated such a possibility for the scalar field cosmology with a minimal and non-minimal coupling to gravity. In the phase space of evolutional scenarios of cosmic evolution pathologies disappear. Trajectories depart from the invariant submanifold $\frac{\dot{H}}{H^2}=0$ of the corresponding dynamical system and this behaviour can be approximated by running cosmological term such as a slow roll parameter $\epsilon_1=\frac{\dot{H}}{H^2} \ll 1$.

\acknowledgments{The work was supported by the grant NCN DEC-2013/09/B/ST2/03455. We are very grateful of A. Krawiec for stimulating discussion and remarks. Especially I would like to thank S.~D. Odintsov and V. Oikonomou for discussion of the problem of a covariance of the vacuum.}

\providecommand{\href}[2]{#2}\begingroup\raggedright\endgroup
\end{document}